\documentstyle[prb,aps,epsf,psfig]{revtex}
\begin{document}
   \oddsidemargin -1.0 cm  \topmargin -0.8 cm
\twocolumn[\hsize\textwidth\columnwidth\hsize
           \csname @twocolumnfalse\endcsname
\title{Electrostatic potential in a superconductor}
\author{Pavel Lipavsk\'y and Jan Kol{\'a}{\v c}ek}
\address{Institute of Physics, Academy of Sciences,
Cukrovarnick\'a 10, 16253 Praha 6, Czech Republic}
\author{Klaus Morawetz}
\address{Max-Planck-Institute for the Physics of Complex
Systems, Noethnitzer Str. 38, 01187 Dresden, Germany}
\author{Ernst Helmut Brandt}
\address{Max-Planck-Institut f\"ur Metallforschung,
         D-70506 Stuttgart, Germany}
\maketitle
\begin{abstract}
The electrostatic potential in a superconductor is studied. 
To this end Bardeen's extension of the Ginzburg-Landau theory 
to low temperatures is used to derive three Ginzburg-Landau 
equations -- the Maxwell equation for the vector potential, 
the Schr\"odinger equation for the wave function and
the Poisson equation for the electrostatic potential.
The electrostatic and the thermodynamic potential
compensate each other to a great extent resulting into 
an effective potential acting on the superconducting condensate.
For the Abrikosov vortex lattice in Niobium, numerical 
solutions are presented and the different contributions to 
the electrostatic potential and the related charge distribution 
are discussed.
\end{abstract}
\vskip2pc]
\section{Introduction}

Even in equilibrium, any inhomogeneous conductor has internal
electric fields which keep its charge distribution close to
local neutrality. The superconductor is not an exception. While the
electrochemical potential is constant, the local chemical
potential varies in general with any gradient in the system. A
distinct property of the superconductor is that in equilibrium
there can be an inhomogeneity due to the diamagnetic electric
current.

The electric field in a superconductor with a stationary
current has been discussed already in 1937 by Bopp\cite{B37}.
 From the hydrodynamic description of a charged liquid, Bopp
has concluded that the inertial and Lorentz force created by
the current are balanced by the Coulomb force. The
corresponding electrostatic potential has the form of a
Bernoulli potential\cite{L50}.

If the Lorentz force dominates, the Bernoulli potential can
also be considered as Hall effect. While it was clear that
there has to be a Hall voltage which passes the Lorentz
force from electrons to the lattice, its measurements by
contacts in standard Hall setups did not show any. It was
understood\cite{H66} that by contacts one observes 
differences in the electrochemical (not electrostatic) 
potential but this potential is constant in equilibrium.

With the aim to distinguish the electrostatic potential from 
the electrochemical one, as late as 1968, Bok and
Klein\cite{BK68} have used the Kelvin capacitive coupling
proposed by Hunt\cite{H66} and have observed first the 
Bernoulli potential on the surface of a superconductor. 
Similar measurements have been performed by Brown and 
Morris\cite{BM68,MB71} or more recently by Chiang and 
Shevchenko\cite{CS86,CS96}.

Even a perfect surface establishes itself a very strong defect 
which essentially modifies the electric field.\cite{LKM01} It is
desirable to observe the internal electric field directly
in the bulk of a superconductor. A new experiment in this
direction has been performed recently by Kumagai {\em et
al\,}\cite{KNM} who have measured the electric field in
a type-II superconductor in mixed state by
nuclear quadrupole resonance.

Another consequence of the electric field in the bulk is
a charge of the vortex core. Blatter {\em et
al\,}\cite{BFGLO96} have proposed an experiment by which
the vortex charge can be accessed. Such measurement,
however, is still to be performed. It is also speculated
that the vortex charge affects the motion of vortices and
thus plays a role in the sign reversal of the Hall
regime\cite{FGLV95}. Since the theory of the anomalous Hall
voltage is still open, one cannot conclude about the
core charge from this effect.

In this paper we derive a phenomenological theory of the
Ginzburg-Landau (GL) type which allows one to evaluate the
electric field in the bulk of superconductors at low
temperatures. A brief presentation of this theory has been
already published in Ref.~\onlinecite{KL01}. Here we
present details and show how to handle numerically this
theory for the Abrikosov lattice of vortices. The
electrostatic potential in the vortex lattice is shown for a
selected temperature and the contribution of the electric
field to forces acting on the condensate is discussed.
Throughout the paper we use the language of the two-fluid
model. The fluid of superconducting electrons is called
condensate while electrons mean normal electrons.

In the next section we review theoretical approaches to the
electric field. In Sec.~\ref{IIIA} we introduce the free energy
which includes the condensation energy of Gorter and Casimir,
the kinetic energy of Ginzburg and Landau, and the standard
electromagnetic energy. Sec.~\ref{IIIB} presents the essential part
of our approach. We use the variational principle to derive
three GL equations: the Maxwell equation for the magnetic
field, the Schr\"odinger equation for the wave function,
and the Poisson equation for the electrostatic potential in
the bulk of superconductors. In Sec.~\ref{IV}, the hydrodynamic
picture is used to link the presented theory with the former
approaches reviewed in Sec.~\ref{II}. In Sec.~\ref{V} we discuss magnetic
properties of the Abrikosov vortex lattice as a function of
the temperature. In Sec.~\ref{VI} we compare the electrostatic
potential with other potentials acting on the condensate.
We also present the charge distribution and show that its
amplitude is very small what allows one to employ a convenient
quasi-neutral approximation. Sec.~\ref{VII} presents the conclusions.
In Appendix~A we estimate the material parameters for
Niobium using the McMillan formula and empirical rules
established from chemical trends.

\section{Historical review} \label{II}

The electric field in superconductors has been studied
since the discovery of superconductivity. Accordingly,
various approaches to this problem can be found in the
literature. We will briefly remind the progress in this field
made mainly in late 1960's and early 1970's.

\subsection{Bernoulli potential} 

The Bernoulli potential for superconductors has been first
derived by Bopp\cite{B37}. Here we follow the later approach
of London\cite{L50}.
The condensate has to obey two equations of motion. First,
it is the London condition,
\begin{equation}   
m{\bf v}=-e{\bf A},
\label{Lc}
\end{equation}
where $\bf v$ is the local velocity of the condensate and $\bf A$
is the vector potential. Second, it is the Newton equation
\begin{equation}   
m\dot{\bf v}=e({\bf E}+{\bf v}\times{\bf B})+{\bf F}_s,
\label{Ne}
\end{equation}
where the first term is the Lorentz force with the electric field,
${\bf E}=-\partial{\bf A} / \partial t -\nabla \varphi$, and the
magnetic field, ${\bf B}=\nabla\times {\bf A}$. The additional
force ${\bf F}_s$ has been treated by different authors within
rather different approximations.

Since the motion of the condensate is fully determined by the
London condition, one can use the Newton equation to
determine the force acting on the condensate. Once the
additional force will be specified, this procedure allows
one to identify the electrostatic potential $\varphi$.

\subsubsection{Time derivative of the London condition} 

To bring the London condition into a form which can be
easily compared with the Newton equation, we take the total
time derivative, $d / dt = \partial / \partial t +
({\bf v}\nabla)$, of the London condition (\ref{Lc}),
\begin{equation}  
m\dot{\bf v}=-e{\partial{\bf A}\over\partial t}-e({\bf v}
\nabla){\bf A}.
\label{Lc1}
\end{equation}
The first term we express via the electric field,
$-\partial{\bf A} / \partial t ={\bf E}+\nabla\varphi$.
For the second term we use a vector identity which in
components reads
\begin{equation}   
{\bf v}_j\nabla_j{\bf A}_i=-[{\bf v}\times\nabla\times{\bf A}
]_i+{\bf v}_j \nabla_i{\bf A}_j.
\label{Lc2}
\end{equation}
In the first term of (\ref{Lc2}) one can recognize the
Lorentz force, $e{\bf v}\times\nabla\times{\bf A}=e{\bf v}
\times{\bf B}$. In the second term of (\ref{Lc2}) we
substitute $\bf A$ by the velocity from the London condition,
$e{\bf v}_j\nabla_i
{\bf A}_j=-m{\bf v}_j\nabla_i{\bf v}_j=-\nabla_i {1\over 2}
mv^2$.

The time derivative of the London condition then reads
\begin{equation} 
m\dot{\bf v}=e({\bf E}+{\bf v}\times{\bf B})+\nabla\left(e
\varphi+{1\over 2}mv^2\right).
\label{Lc3}
\end{equation}
This equation can be compared with the Newton equation
(\ref{Ne}) giving the electrostatic potential as
\begin{equation} 
\nabla e\varphi={\bf F}_s-\nabla{1\over 2}mv^2.
\label{Lc4}
\end{equation}

\subsubsection{Bernoulli potential}  

London assumed that the motion of the condensate is controlled
by the Lorentz force only. In this approximation, there is
no additional force,
\begin{equation} 
{\bf F}_s=0.
\label{af1}
\end{equation}
 From (\ref{Lc4}) thus follows the electrostatic potential
of the Bernoulli type,
\begin{equation} 
e\varphi=-{1\over 2}mv^2.
\label{Bp}
\end{equation}

\subsubsection{Quasiparticle screening}  

In 1964 van~Vijfeijken and Staas\cite{VS64} have extended the
Bernoulli potential to finite temperatures using the two fluid
model.
When flowing, normal electrons dissipate energy. Therefore, in
the stationary case they have to stay at rest in spite of
the presence of an electric field. These authors have
introduced an unspecified force,
\begin{equation}    
{\bf F}_n=e\nabla\varphi,
\label{sVn}
\end{equation}
acting on
electrons to keep them at rest, ${\bf F}_n+e{\bf E}=0$.
This force is assumed to result from the interaction between
the electrons and the condensate. Accordingly, there has to
be a reaction force ${\bf F}_s$ acting on the condensate so
that the Newton law of action and reaction is fulfilled,
\begin{equation}     
n_n{\bf F}_n+n_s{\bf F}_s=0,
\label{sVar}
\end{equation}
where $n_n$ and $n_s$ are densities of electrons and condensate.
  From (\ref{sVn}) and (\ref{sVar}) one finds the additional
force,
\begin{equation} 
{\bf F}_s=-{n_n\over n_s}e\nabla\varphi,
\label{sVs}
\end{equation}
and from (\ref{Lc4}) follows the electrostatic potential
\begin{equation} 
e\varphi=-{n_s\over n}{1\over 2}mv^2.
\label{vS}
\end{equation}
This is the Bernoulli potential (\ref{Bp}) reduced by the
share of the condensate on the total density, $n=n_n+n_s$.

The reduction of the Bernoulli potential has become known as
``screening by normal electrons'' or ``quasiparticle screening''.
The quasiparticle screening, however, has to be distinguished
from the Thomas-Fermi screening present in all metals
including superconductors.

\subsubsection{Thomas-Fermi screening}  

In superconductors, the screening is the same as in normal
metals. Starting from  the time-dependent Ginzburg-Landau
theory, Jakeman and Pike\cite{JP67} have derived the Poisson
equation for the electric field with the reduced Bernoulli
potential as the driving term,
\begin{equation}      
e\varphi-\lambda_{TF}^2\nabla^2e\varphi
=-{n_s\over n}{1\over 2}mv^2.
\label{phiscr}
\end{equation}

Currents change typically on the scale of the London
penetration depth or the GL coherence length, which are much
larger than the Thomas-Fermi screening length $\lambda_{TF}$.
The electrostatic potential $\varphi$ thus can be treated in
the limit of strong screening, $\lambda_{TF}\to 0$, and
from (\ref{phiscr}) one recovers (\ref{vS}).

\subsubsection{Thermodynamic potential}

Already in 1949 Sorokin\cite{S49} has followed the
hydrodynamic approach of Bopp assuming an unspecified free
energy,
\begin{equation}   
{\cal F}_s=\int d{\bf r} f_s,
\label{Sorok1}
\end{equation}
responsible for the superconducting transition. Here $f_s$
is the density of free energy and $d{\bf r}$ denotes
integration over the sample volume.
The free energy leads to a thermodynamic potential,
\begin{equation}    
w_s={\delta{\cal F}_s\over\delta n_s}
={\partial f_s\over\partial n_s},
\label{Sorokchi}
\end{equation}
which yields the additional force
\begin{equation} 
{\bf F}_s=-\nabla w_s.
\label{Ne1}
\end{equation}
According to (\ref{Lc4}) the Bernoulli potential is modified
as
\begin{equation}    
e\varphi=-{1\over 2}mv^2-w_s.
\label{BpS}
\end{equation}
The quasiparticle screening is one of the contributions that
result from the thermodynamic potential. There are also
other contributions which can provide information about
the pairing mechanism.

Unfortunately, London has disregarded the thermodynamic
potential in his book\cite{L50} as unknown and unimportant.
His objection was correct at that time since the first reliable
thermodynamic potential has been derived eight years
later by Bardeen, Cooper and Schrieffer\cite{BCS57}.
On the other hand, the two-fluid free energy of Gorter and
Casimir\cite{GC34a,GC34b,GC34c} known from 1934, could be
used within Sorokin's approach to provide at least
qualitative results. Our approach follows Sorokin, except
that we use an explicit thermodynamic potential of Gorter
and Casimir and a non-local kinetic energy.

\subsubsection{Non-local corrections}  
As shown in Ref.~\onlinecite{KLB01}, London's approach can
be modified towards strongly inhomogeneous systems using the
Schr\"odinger equation for a Cooper pair,
\begin{equation}   
{1\over 2m^*}\left(-i\hbar\nabla-e^*{\bf A}\right)^2\psi+
e^*\varphi\psi+2w_s\psi=0,
\label{Se1}
\end{equation}
instead of the Newton equation (\ref{Ne}). Here we have
also included the thermodynamic potential $w_s$ neglected
in Ref.~\onlinecite{KLB01}.

 From (\ref{Se1}) follows directly a quantum modification
of the Bernoulli potential,
\begin{equation}    
e^*\varphi=-{1\over\psi}{1\over 2m^*}(-i\hbar\nabla-
e^*{\bf A})^2\psi-2w_s.
\label{nBp}
\end{equation}
In the quasi-classical approximation, $(-i\hbar\nabla-
e^*{\bf A})\psi=m^*{\bf v}\psi$, this formula reduces to
potential (\ref{BpS}) derived by Sorokin.

To obtain the actual value of the potential, the wave
function $\psi$ is identified with the GL wave
function and solved from the GL equation. Accordingly,
the Cooperon mass and charge, $m^*=2m$ and $e^*=2e$,
appear in the Schr\"odinger equation (\ref{Se1}).

\subsection{Thermodynamic correction} 

Rickayzen\cite{R69} proposed a thermodynamic approach to the
electric field. He assumes a quadratic dependence of the
free energy on the velocity, what limits his study to weak
currents. For systems with a parabolic band, the
increase of the free energy due to the current equals the
kinetic energy of the condensate,
\begin{equation}    
f_{\rm kin}=n_s{1\over 2}mv^2.
\label{Ric1}
\end{equation}

The electrochemical potential, $\nu=E_F+
\nu_{\rm kin}+e\varphi$, is constant in the whole system,
therefore $e\varphi=-\nu_{\rm kin}$. Since $\nu= \partial
 f / \partial n$, the velocity variation of the local
chemical potential is $\nu_{\rm kin}= \partial f_{\rm kin}
/ \partial n$. Accordingly, the electrostatic potential
induced by the current reads\cite{R69}
\begin{equation}    
e\varphi=-{\partial n_s\over\partial n}{1\over 2}mv^2.
\label{Ric2}
\end{equation}

Expression (\ref{Ric2}) generalizes (\ref{vS}). From the
phenomenological density of the condensate,
\begin{equation}     
n_s=n\left(1-{T^4\over T_c^4}\right),
\label{Ric3}
\end{equation}
follows
\begin{equation}     
e\varphi=-{n_s\over n}{1\over 2}mv^2+
4{n_n\over n}{\partial\ln T_c\over\partial\ln n}
{1\over 2}mv^2.
\label{Ric4}
\end{equation}
The first term is the reduced Bernoulli potential (\ref{vS}),
the second is a thermodynamic correction. According to
(\ref{Ric3}), the first term of (\ref{Ric4}) depends on
the temperature as $1-T^4/T_c^4$ while the second one goes as
$T^4/T_c^4$. At higher temperatures the second term
dominates.

\subsubsection{BCS estimate}  

The density dependence of $T_c$ reflects the pairing
mechanism. Its magnitude can be estimated from the BCS
relation\cite{BCS57},
\begin{equation}   
k_BT_c=1.14\hbar\omega_D{\rm e}^{-{1\over{\cal D}V}},
\label{BCSTc}
\end{equation}
where ${\cal D}$ is the single-spin density of states,
$\omega_D$ is the cut-off frequency usually approximated
by the Debye temperature, $\hbar\omega_D\approx k_B\theta_D$,
and $V$ is the BCS interaction. Assuming that $\theta_D$ and $V$
do not depend on the density, one finds
\begin{equation} 
{\partial\ln T_c\over\partial\ln n}=
{\partial{\cal D}\over\partial\ln n} {1\over{\cal D}^2V}
\approx-
{\partial\ln{\cal D}\over\partial\ln n}\ln {\theta_D\over T_c}.
\label{BCSTder}
\end{equation}

It remains to estimate the derivative of the density of states.
For systems with a parabolic band the density of states
is proportional to the Fermi momentum, $\ln{\cal D}\propto
k_F$, while the density of electrons is $n \propto k_F^3$.
Accordingly, $\partial\ln{\cal D}/\partial\ln n\approx 1/3$. 
For Niobium we have a very similar value $\partial\ln{\cal D}/ 
\partial\ln n=0.32$, see Tab.~1 in Appendix~A.

With the BCS estimate (\ref{BCSTder}), the electrostatic
potential (\ref{Ric4}) reads
\begin{equation}  
e\varphi=-{1\over 2}mv^2\left(
{n_s\over n}+
{n_n\over n}{4\over 3}\ln{\theta_D\over T_c}\right).
\label{Dderest}
\end{equation}
For conventional superconductors, $\theta_D/T_c$ is of the order
of few tens, therefore the thermodynamic correction is the
dominant contribution for approximately $T > {2\over 3}T_c$.

For Niobium the BCS formula (\ref{Dderest}) overestimates
the thermodynamic correction. The approximate factor from
(\ref{Dderest}) is $(4/3)\ln(\theta_D/T_c)=4.5$ while the full
factor from (\ref{Ric4}) gives $-4(\partial\ln T_c/\partial\ln
n)=3.0$, see (\ref{B10}) and Appendix~A.

\subsubsection{BCS microscopic theory}  

Within the BCS theory, the electric field has been studied
by Adkins and Waldram\cite{AW68}, Rickayzen\cite{R69} and
Hong\cite{H75}. In all these studies, materials with a
general band structure have been addressed. For the sake
of simplicity we discuss only the parabolic band, for which
the BCS theory yields\cite{AW68,H75}
\begin{equation}     
e\varphi\approx{\Delta^2\over 2}
{\partial\ln{\cal D}\over\partial E_F}
\ln\left({2\omega_D\hbar\over\Delta_0}\right).
\label{phiT}
\end{equation}
Here $\Delta_0=1.75 k_BT_c$ is the gap at $T=0$ and
$\Delta$ is the actual local value of the gap.

Since the electric current locally depresses the gap,
$\Delta=\Delta_{\rm eq}+\Delta'$ with $\Delta'\propto-
v^2$, the potential (\ref{phiT}) includes the
contribution of Bernoulli type, $\varphi=\varphi_{\rm eq}+
\varphi'$ with $\varphi'\propto -v^2$. As shown by
Rickayzen\cite{R69}, $\varphi'$ can be rearranged into
the thermodynamic correction of (\ref{Dderest}).

\subsection{Aims of the present approach}  

In this paper we discuss the Ginzburg-Landau theory
modified in two directions. First, following Bardeen we
use its extension to low temperatures. Second, we include
the electrostatic potential. We focus on the bulk of
superconductors, i.e., on regions which are far from the
surface on the scale of Thomas-Fermi screening length.

Starting from the free energy, we derive the Poisson equation
along with the Maxwell equation for the vector potential
and the equation of the Schr\"odinger type for the wave
function. The presented theory yields
\begin{itemize}
\item
non-local Bernoulli potential,
\item
quasiparticle screening,
\item
thermodynamic corrections,
\item
thermoelectric field of normal metal at $T=T_c$,
\item
Thomas-Fermi screening.
\end{itemize}

Our approach parallels the original study of Sorokin, however,
we use the explicit phenomenological free energy proposed by
Bardeen. It combines the Ginzburg-Landau (GL) theory with the
Gorter-Casimir free energy. Naturally, this theory is only
approximate. Its major advantage is its transparency and a
simple implementation scheme.

\section{Extended Ginzburg-Landau theory}  \label{III}

\subsection{Free energy}\label{IIIA}

Bardeen\cite{B55,B54} has extended the GL theory\cite{GL50,T96}
by the use of the Gorter-Casimir two-fluid
model\cite{GC34a,GC34b,GC34c} so as to apply to all
temperatures. We briefly recall the Gorter-Casimir model
and introduce other components of the free energy.

\subsubsection{Condensation energy of two-fluid model}  

Gorter and Casimir assumed that the superconducting state is
characterized by an order parameter $\varpi$ which is zero in
the normal state and unity at zero temperature. They have
modified the normal state density of free energy as
\begin{equation}   
f_s=U-\varepsilon_{\rm con}\varpi-
{1\over 2}\gamma T^2\sqrt{1-\varpi}.
\label{GCFp}
\end{equation}
For $\varpi=0$, the free energy (\ref{GCFp}) equals the
normal state free energy consisting of the internal energy $U$
and the entropy term $-{1\over 2}\gamma T^2$. Sommerfeld's
$\gamma$ is the linear coefficient of the specific heat.
In the superconducting state, $\varpi\ne 0$, two mechanisms
are expected. First, the ordering releases the condensation
energy $\varepsilon_{\rm con}\varpi$. Second, the ordering
reduces the entropy by the factor $\sqrt{1-\varpi}$.

In equilibrium the free energy reaches its minimum. From
$\delta{\cal F}_s / \delta\varpi = \partial f_s / \partial
\varpi =0$ follows that the equilibrium value of the order
parameter is a solution of
\begin{equation}   
\varepsilon_{\rm con}={\gamma T^2\over 4\sqrt{1-\varpi}}.
\label{GCFc}
\end{equation}
At the critical temperature the ordering vanishes, $\varpi=0$,
therefore
\begin{equation}    
\varepsilon_{\rm con}={1\over 4}\gamma T_c^2.
\label{EconTc}
\end{equation}

  From (\ref{GCFc}) with (\ref{EconTc}) follows
\begin{equation}    
\varpi=1-{T^4\over T_c^4},
\label{omT}
\end{equation}
which agrees with the observed temperature dependence of
the condensate density (\ref{Ric3}). Accordingly, one can
identify the order parameter $\varpi$ with the fraction
of the condensate on the total density of electrons,
\begin{equation}    
\varpi={n_s\over n}.
\label{omshare}
\end{equation}

\subsubsection{Kinetic energy}  

An electric current contributes to the free energy by
the kinetic energy of the condensate. The kinetic energy
proposed by Ginzburg and Landau\cite{GL50,T96} reads
\begin{equation}  
{\cal F}_{\rm kin}=\int d{\bf r}{1\over 2m^*}
\left|\left(-i\hbar\nabla-e^*{\bf A}\right)\psi\right|^2,
\label{FK}
\end{equation}
with the wave function normalized as
\begin{equation}  
|\psi|^2={n_s\over 2}.
\label{normpsi}
\end{equation}
We use the isotropic effective mass for simplicity, the
anisotropic case will be discussed in the next paper.

In the Gorter-Casimir free energy we thus substitute
the order parameter by
\begin{equation}  
\varpi={2|\psi|^2\over n}={2|\psi|^2\over 2|\psi|^2+n_n}.
\label{ompsi}
\end{equation}

\subsubsection{Electromagnetic energy} 

The density of free energy has four components,
\begin{equation}   
{\cal F}={\cal F}_s+{\cal F}_{\rm kin}+{\cal F}_C+{\cal F}_M.
\label{FGL}
\end{equation}
The free energy ${\cal F}_s$ given by the volume integral 
(\ref{Sorok1}) of the free energy density (\ref{GCFp}), we 
shall call the condensation energy according
to its most important part. The kinetic energy,
${\cal F}_{\rm kin}$, is given by the GL expression
(\ref{FK}).
The Coulomb interaction reads
\begin{equation}   
{\cal F}_C={1\over 2}\int\!\!\!\!\int d{\bf r}d{\bf r}'
{1\over 4\pi\epsilon}{1\over|{\bf r}-{\bf r}'|}
\rho({\bf r})\rho({\bf r}'),
\label{Coulsep}
\end{equation}
where $\rho=e^*|\psi|^2+en_n+\rho_{\rm latt}$ is the charge
density. The Coulomb interaction also determines the
electrostatic potential, by
\begin{equation}   
\varphi({\bf r})=\int d{\bf r}'{1\over 4\pi\epsilon}
{1\over|{\bf r}-{\bf r}'|}\rho({\bf r}')
\label{Coulsep2d}
\end{equation}
or in its differential form by the Poisson equation
\begin{equation}   
-\epsilon\nabla^2\varphi=\rho.
\label{Pe}
\end{equation}

Finally, the Helmholtz magnetic free energy reads\cite{W96},
\begin{equation}    
{\cal F}_M=\int d{\bf r}{1\over 2\mu_0}({\bf B}-{\bf B}_a)^2,
\label{Fm}
\end{equation}
where ${\bf B}_a$ is the applied magnetic field.

The total free energy is a functional of the wave function,
the vector potential and the (normal) electron density,
${\cal F}[\psi,{\bf A},n_n]$. The other physical quantities
like $\bf B$, $\varphi$, $n$, $n_s$, $\rho$ or $\varpi$ are
subsidiary and have to be understood as functions of the
independent variables $\psi$, $\bf A$ and $n_n$.

We note that much more sophisticated approximations of the
free energy ${\cal F}_s+{\cal F}_{\rm kin}$ have been
developed from the BCS theory and Eliashberg's theory
already in 1960's, see e.g. Ref.~\onlinecite{W69}. In
principle, one can start from any of these approximations.
Since our prime interest is in the electrostatic potential
and the related charge distribution, we prefer to use the
simple approximation of Bardeen.

\subsection{Ginzburg-Landau equations of motions}\label{IIIB}

In equilibrium, the system stays in the state with minimum
free energy. Accordingly, the variations of ${\cal F}$ with
respect to the vector potential $\bf A$, the wave function 
$\psi$ and the electron density $n_n$ have to vanish.

During the variation procedure, the two-point function 
${\cal F}_C$ and the one-point functions (all the others) 
are treated differently. The local contributions are given 
by the corresponding densities, $f_\alpha[l(r),\nabla l(r)]$, 
with ${\cal F}_\alpha=\int d{\bf r}f_\alpha$, and their 
variation is of Lagrange's form\cite{T61},
\begin{equation}    
{\delta{\cal F}_\alpha\over\delta l}=
{\partial f_\alpha\over\partial l}-
\nabla{\partial f_\alpha\over\partial\nabla l}.
\label{varpar}
\end{equation}
Here $\alpha$ represents the subscripts $s$, kin and $M$ while $l$
stands for $\bf A$, $\psi$ or $n_n$.

The variation of the Coulomb energy with respect to $\psi$
or $n_n$ can be expressed by the variation with
respect to the density of charge $\rho$ which reads
\begin{equation}  
{\delta{\cal F}_C\over\delta\rho({\bf r})}=
\int d{\bf r}'{1\over 4\pi\epsilon}
{1\over|{\bf r}-{\bf r}'|}\rho({\bf r}').
\label{Coulsep2}
\end{equation}
According to (\ref{Coulsep2d}) we can abbreviate this
variation as
\begin{equation}  
{\delta{\cal F}_C\over\delta\rho}=\varphi.
\label{Coulsep2p}
\end{equation}

\subsubsection{Maxwell equation}

The vector potential $\bf A$ appears in the kinetic energy
${\cal F}_{\rm kin}$, and its gradients enter the magnetic free
energy via ${\bf B}=\nabla\times{\bf A}$. From the condition
of minimum with respect to $\bf A$, one recovers the Maxwell
equation,
\begin{equation}  
\nabla\times\nabla\times{\bf A}=\mu_0{\bf j},
\label{Me}
\end{equation}
where the current $\bf j$ is given by the quantum-mechanical
formula\cite{T96},
\begin{equation}   
{\bf j}={e^*\over m^*}{\rm Re}\,\bar\psi(-i\hbar\nabla-e^*
{\bf A})\psi,
\label{Mej}
\end{equation}
known as the second GL equation. Here $\bar\psi$ denotes the
complex conjugate of $\psi$.

\subsubsection{Schr\"odinger equation} 

The wave function $\psi$ enters the free energy via the order
parameter, the Coulomb interaction via the charge density,
and the kinetic energy ${\cal F}_{\rm kin}$, where also the
gradients of $\psi$ appear. The variation parallels the
derivation of the first GL equation\cite{T96}, for details
see Ref.~\onlinecite{W96}.

The variation with respect to $\bar\psi$ leads to the equation
of the Schr\"odinger type,\footnote{
The free energy is a real function which depends on the
complex function $\psi$ and its conjugate $\bar\psi$. We
express absolute values as products, $|\psi|^2=\bar\psi\psi$,
and take $\delta\psi$ and $\delta\bar\psi$ as independent
perturbations, $\delta{\cal F}= (\delta{\cal F} / \delta
\bar\psi) \delta\bar\psi + (\delta{\cal F} / \delta\psi )
\delta\psi$. Since $\delta{\cal F} / \delta\bar\psi =
\overline{\delta{\cal F}} / \delta\psi $, both variations
vanish at the same time. The variational condition
$ \delta{\cal F} / \delta\bar\psi =0$ yields the
Schr\"odinger equation.}
\begin{equation}   
{1\over 2m^*}(-i\hbar\nabla-e^*{\bf A})^2\psi+\chi\psi=0.
\label{Se}
\end{equation}
The effective potential,
\begin{equation}   
\chi={\delta\over\delta|\psi|^2}\left({\cal F}_C+{\cal F}_s
\right)=e^*\varphi+{\partial f_s\over\partial|\psi|^2},
\label{Sechi}
\end{equation}
covers all forces acting on Cooper pairs.

\subsubsection{Diffusion of normal electrons}  

 From the variation with respect to the electron density,
$\delta{\cal F} / \delta n_n =0$, one finds that the sum
of all potentials acting on the normal electrons has to vanish,
i.e.,
\begin{equation}   
e\varphi=-{\delta{\cal F}_s\over\delta n_n}=-
{\partial f_s\over\partial n_n}.
\label{nvar}
\end{equation}
This condition parallels Eq.~(\ref{sVn}) of van~Vijfeijken and
Staas.

The set of equations (\ref{Me}-\ref{nvar}) is closed. Its
particular form is given by the condensation energy $f_s$.
Below we evaluate derivatives of the condensation energy
within the Gorter-Casimir approximation (\ref{GCFp}).

\subsubsection{Effective potential acting on Cooper pairs}  

To describe the motion of the condensate given by the
effective potential $\chi$, we have to evaluate the electrostatic
potential. This will be done in the spirit of van~Vijfeijken
and Staas using the equation for normal electrons (\ref{nvar}).

Since $e^*=2e$, the effective potential results
from (\ref{nvar}) and (\ref{Sechi}) as
\begin{equation}  
\chi={\partial f_s\over\partial|\psi|^2}-2
{\partial f_s\over\partial n_n}.
\label{chi1}
\end{equation}

The combination of the partial derivatives excludes a
contribution of functions which depend exclusively on the
total density, $\partial n / \partial|\psi|^2 -
2 \partial n / \partial n_n =0$. Accordingly, derivatives of
density-dependent material parameters ($U$, $\varepsilon_{\rm
con}$, $\gamma$ and $T_c$) do not contribute to the potential
$\chi$. From (\ref{chi1}) and (\ref{GCFp}) thus follows
\begin{equation}  
\chi=-2{\varepsilon_{\rm con}\over n}+{\gamma T^2\over 2n}
{1\over\sqrt{1-{2|\psi|^2\over n}}}.
\label{chi2}
\end{equation}
With the potential (\ref{chi2}), the Schr\"odinger equation
(\ref{Se}) is identical to the extended GL equation
proposed by Bardeen\cite{B55,B54}.

Close to $T_c$, the potential $\chi$ approaches the
quadratic form of Ginzburg and Landau,
\begin{equation}   
\chi\to\alpha+\beta|\psi|^2,
\label{chi3}
\end{equation}
with
\begin{equation}  
\alpha={\gamma T_c\over n}(T-T_c),~~~~~~~~
\beta={\gamma T_c^2\over 2n^2}.
\label{chi4}
\end{equation}

We note that the effective potential $\chi$ depends
on the density $n$ via $\gamma$ and $T_c$. In principle, one
has to iterate the GL equation together with relations for the
density $n$. In practice, deviations of the density from its
crystal value are very small, $|n+\rho_{\rm latt}/e|\ll n$,
and the approximation $en\approx-\rho_{\rm latt}$ is well
justified when one solves for $\psi$ and $\bf A$.

\subsubsection{Poisson equation with screening}  

Now we rearrange (\ref{nvar}) into the form of the Poisson
equation with the Thomas-Fermi screening. The variation of
the free energy in (\ref{nvar}) reads
\begin{equation}  
{\partial f_s\over\partial n_n}={\partial U\over\partial n}-
{\partial\over\partial n_n}\left(\varepsilon_{\rm con}\varpi+
{1\over 2}\gamma T^2\sqrt{1-\varpi}\right).
\label{U1}
\end{equation}

The derivative of the internal energy,
\begin{equation}   
{\partial U\over\partial n}=E_F,
\label{U2}
\end{equation}
is the Fermi energy at the normal ground state
of total electron density $n$. In the presence of the
electrostatic potential, the density $n$ differs from its
crystal value, $n_0=-\rho_{\rm latt} / e $, by the
density perturbation $\rho / e$. The Fermi energy thus
depends on the charge density as
\begin{equation}   
E_F=E_F^0+
{\partial E_F\over\partial n}{\rho\over e},
\label{U3}
\end{equation}
where $E_F^0$ is the crystal value. As usual in the
theory of superconductivity, we associate the crystal Fermi
energy with the origin of the energy scale,
$E_F^0=0$.

The density dependence of the local Fermi energy determines
the screening. The density derivative of the Fermi energy is
the inverse density of states,
\begin{equation}  
{\partial E_F\over\partial n}={1\over 2{\cal D}}.
\label{U4}
\end{equation}
Using the Poisson equation (\ref{Pe}) to evaluate the charge
from the electrostatic potential, $\rho=-\epsilon\nabla^2
\varphi$, one can express the Fermi energy as
\begin{equation}  
E_F=-\lambda_{TF}^2\nabla^2e\varphi.
\label{U5}
\end{equation}
with the Thomas-Fermi screening length $\lambda_{TF}^2=
\epsilon / ( 2{\cal D}e^2 )$.

By a substitution of the Fermi energy (\ref{U5}) in the
stability condition for normal electrons (\ref{nvar}), we
arrive at the screened Poisson equation,
\begin{equation} 
e\varphi-\lambda_{TF}^2\nabla^2e\varphi=
{\partial\over\partial n_n}\left(\varepsilon_{\rm con}\varpi+
{1\over 2}\gamma T^2\sqrt{1-\varpi}\right) .
\label{chi5}
\end{equation}
The right hand side of (\ref{chi5}) is readily evaluated
from the assumption that the condensation energy
$\varepsilon_{\rm con}$ and the Sommerfeld $\gamma$ depend
only on the total density $n=2|\psi|^2+n_n$, and from the
explicit form of the order parameter,
$\varpi= 2|\psi|^2 / ( 2|\psi|^2+n_n )$, giving
\begin{eqnarray} 
e\varphi&-&\lambda_{TF}^2\nabla^2e\varphi
\nonumber\\
&=&\chi{|\psi|^2\over n}+
{\partial\varepsilon_{\rm con}\over\partial n}
{2|\psi|^2\over n}+{T^2\over 2}{\partial\gamma\over\partial n}
\sqrt{1-{2|\psi|^2\over n}}.
\label{chi6}
\end{eqnarray}
In the language of Jakeman and Pike, the Poisson equation
(\ref{chi6}) is called third GL equation.

The first term on the rhs of (\ref{chi6}) is the
non-local Bernoulli potential with quasiparticle screening.
This can be seen if we multiply the GL equation (\ref{Se}) by
$\bar\psi$ which  yields
\begin{equation}  
\chi{|\psi|^2\over n}=-{1\over 2m^*n}
\bar\psi\left(-i\hbar\nabla-e^*{\bf A}\right)^2\psi.
\label{chi7}
\end{equation}
In the classical approximation of the kinetic energy,
$(1 / 2m^*)\bar\psi(-i\hbar\nabla-e^*{\bf A} )^2\psi\approx
{1\over 2}m^*v^2|\psi|^2$, one finds that the first term of
(\ref{chi6}) is the screened Bernoulli potential
of van~Vijfeijken and Staas,
$\chi|\psi|^2/n\approx -(n_s / n){1\over 2}m v^2$.

The second and third terms of (\ref{chi6}) are non-linear
generalizations of the thermodynamic correction by Rickayzen.
Note that the third term remains finite at the critical
point, $T\to T_c$ and $|\psi|\to 0$, yielding the normal state
thermoelectric field\cite{Z60}.

The set of GL equations is closed. It consists of the Maxwell
equation (\ref{Me}) with the current (\ref{Mej}), the
Schr\"odinger equation (\ref{Se}) with the potential
(\ref{chi2}), and the screened Poisson equation (\ref{chi6}).
Deviations from the local charge neutrality are given by the
bare Poisson equation (\ref{Pe}).

\section{Hydrodynamic picture} \label{IV}

Within the thermodynamic approach of Sec.~\ref{IIIB}, the
electrostatic potential $\varphi$ is a function of the
wave function $\psi$. This contrasts with the original
derivations of the Bernoulli potential expressed in terms of
the condensate velocity $\bf v$. To make the link with the
original approaches mentioned in Sec.~\ref{II}, in this section
we reformulate the above thermodynamic theory in the
hydrodynamic picture.

The hydrodynamic picture is readily obtained writing the
wave function in terms of the condensate density
(\ref{normpsi}) and the phase $\theta$,
\begin{equation}  
\psi=\sqrt{{n_s\over 2}}\ {\rm e}^{i\theta}.
\label{B1}
\end{equation}
A velocity defined via the current, ${\bf j}=en_s{\bf v}$,
then reads
\begin{equation}  
{\bf v}={1\over m^*}\left(\hbar\nabla\theta-e^*{\bf A}\right).
\label{B2}
\end{equation}

\subsection{On Sorokin's relation}  
In the representation (\ref{B1}) the Schr\"odinger equation
(\ref{Se}) reads\footnote{Equation (\ref{B3}) is the
energy-conserving integral of motion of the Newton-like form
of the Schr\"odinger equation. This Newton-like equation itself
may be found e.g.  in The Feynman Lectures on Physics\cite{F65}.}
\begin{equation}  
{1\over 2}mv^2-{\hbar^2\over 8m}{1\over\sqrt{n_s}}\nabla^2
\sqrt{n_s}+e\varphi+w_s=0,
\label{B3}
\end{equation}
where we have used $m^*=2m$ and $\chi=2e\varphi+2w_s$ as it
follows from (\ref{Sechi}). The thermodynamic potential $w_s$ 
is given by
\begin{equation}  
w_s={1\over 2}{\partial f_s\over\partial|\psi|^2},
\label{B4}
\end{equation}
which is equivalent to (\ref{Sorokchi}). With the non-local
correction neglected, $\nabla^2\sqrt{n_s}\approx 0$, equation
(\ref{B3}) turns into the Sorokin result (\ref{BpS}).

Naturally, the explicit evaluation of the electrostatic
potential within the Sorokin approach parallels the non-local
approach presented in the previous section. Since this approach
is quite transparent, we show this procedure in detail.

Relation (\ref{chi2}) reads
\begin{equation}  
e\varphi+w_s=-{\varepsilon_{\rm con}\over n}+
{\gamma T^2\over 4n}{1\over\sqrt{1-{n_s\over n}}}.
\label{B5}
\end{equation}
Substituting (\ref{B5}) into the local approximation of
(\ref{B3}) one finds
\begin{equation}  
{1\over 2}mv^2={\varepsilon_{\rm con}\over n}-
{\gamma T^2\over 4n}{1\over\sqrt{1-{n_s\over n}}},
\label{B6}
\end{equation}
which yields the condensate density as a function of the
local velocity. Provided that the profile of velocities in the
system is known, from (\ref{B6}) and (\ref{chi6}) one can 
directly evaluate the electrostatic potential.

\subsection{On Rickayzen's result} 
Now we recover Rickayzen's result (\ref{Ric4}). To this end
we have to accept identical approximations. First we neglect
the Thomas-Fermi screening, so that (\ref{chi6}) reads
\begin{equation}  
e\varphi={n_s\over n}(e\varphi+w_s)+{n_s\over n}
{\partial\varepsilon_{\rm con}\over\partial n}+
{T^2\over 2}{\partial\gamma\over\partial n}
\sqrt{1-{n_s\over n}}.
\label{B7}
\end{equation}

Second, Rickayzen assumes a local relation between the
velocity and the electrostatic potential. Accordingly, we
neglect the gradient correction in (\ref{B3}), i.e., we use
the Sorokin approximation, $e\varphi+w_s=-(1/2)mv^2$, with 
the help of which we eliminate the Sorokin potential $w_s$ 
from (\ref{B7}). 

Third, following Rickayzen we take the limit of weak currents,
$n\, mv^2/2\ll\varepsilon_{\rm con}-\gamma T^2/4$. Up to linear
orders in the kinetic energy, from (\ref{B6}) follows $n_s=
n_s^0+n_s'$ with
\begin{equation}
n_s'=-n{T^4\over T_c^4}{n\over\varepsilon_{\rm con}}mv^2,
\label{nsprime}
\end{equation}
so that from (\ref{B7}) results the electrostatic potential
$\varphi=\varphi_{\rm eq}+\varphi'$ as
\begin{eqnarray}  
e\varphi'=&-&{n_s\over n}{1\over 2}mv^2
\nonumber\\
&-&{\partial\varepsilon_{\rm con}\over\partial n}
{n_n\over\varepsilon_{\rm con}}mv^2 +
{T^2\over 2}{\partial\gamma\over\partial n}
{1\over 8}{n\gamma T^2\over\varepsilon_{\rm con}^2}mv^2.
\label{B10}
\end{eqnarray}
Using (\ref{EconTc}) and (\ref{Ric3}),
one can rearrange expression (\ref{B10}) into the potential
(\ref{Ric4}) derived by Rickayzen.

In summary, the result of Sorokin corresponds to the local
approximation of the presented approach. The potential of
Bernoulli type derived by Rickayzen includes further
approximations, in particular the limit of the weak electric
current. We note that for systems with vortices the non-local 
approach is necessary since the ``classical'' kinetic energy
$mv^2/2$ diverges at the vortex center. This divergence is
compensated by the non-local correction so that the
``quantum'' kinetic energy remains regular.

\section{Magnetic properties of the Abrikosov vortex lattice} \label{V}

  In this section we evaluate the wave function and the
magnetic field for the Abrikosov vortex lattice in Niobium.
Pure Niobium is close to the border between type-I and type-II
superconductors since its GL parameter $\kappa=0.78$ is
only slightly above $1/\sqrt{2}$. However, the GL parameter can
be increased up to about three by impurities. For simplicity we
neglect the effect of impurities on material parameters other
than the GL parameter $\kappa$.

As will be proven in the next section, deviations of the total
density of electrons from its unperturbed value are very small,
$|\rho|\ll\rho_{\rm latt}$. We will neglect these deviations
and treat the material parameters in the approximation of
quasi-neutrality, $\gamma(n)\approx \gamma(n_0)$ etc. In this
approximation, the first and the second GL equations are
independent of the third GL equation. Therefore, we shall ignore
the electrostatic potential and related charge deviation
within this section.

\subsection{Dimensionless notation} 
Our approach parallels Ref.~\onlinecite{B97}. In our
calculations we shall use dimensionless quantities,
\begin{eqnarray} 
t&=&{T\over T_c},
\nonumber\\
{\bf b}&=&{\lambda_{\rm Lon}{\bf B}\over\lambda_0\sqrt{B_cB_0}},
\nonumber\\
{\bf a}&=&{{\bf A}\over\lambda_0\sqrt{B_cB_0}},
\nonumber\\
\tilde{\bf r}&=&{{\bf r}\over\lambda_{\rm Lon}}.
\label{redquant}
\end{eqnarray}
Close to the critical temperature, $t\to 1$, these dimensionless
variables reduce to the usual form. \cite{W96}

The thermodynamical critical field $B_c$, the London penetration
depth $\lambda_{\rm Lon}$, and GL parameter $\kappa$ depend on
the temperature as
\begin{eqnarray} 
B_c(t)&=&B_0(1-t^2),
\nonumber\\
\lambda_{\rm Lon}(t)&=&{\lambda_0\over\sqrt{1-t^4}},
\nonumber\\
\kappa(t)&=&\kappa_0\sqrt{{2\over 1+t^2}},
\nonumber\\
 B_{c2}(t) &=& \sqrt{2}\kappa B_c
       = 2B_0 \kappa_0 {1-t^2 \over\sqrt{1+t^2}}.
\label{critval}
\end{eqnarray}
The asymptotic values of these quantities in terms of the
parameters of the Gorter-Casimir model read
\begin{eqnarray} 
B_0&=&T_c\sqrt{{\mu_0\gamma\over 2}},
\nonumber\\
\lambda_0&=&\sqrt{{m\over e^2n\mu_0}},
\nonumber\\
\kappa_0&=&{mT_c\over ne\hbar}\sqrt{{\gamma\over\mu_0}}.
\label{kap0}
\end{eqnarray}

Finally, we introduce a dimensionless amplitude of the
wave function and the dimensionless velocity,
\begin{eqnarray} 
\omega&=&{2|\psi|^2\over n(1-t^4)},
\nonumber\\
{\bf Q}&=&{\bf a}-{1\over\kappa}\tilde\nabla\theta.
\label{psired}
\end{eqnarray}
Our dimensionless notation is identical to
Ref.~\onlinecite{B97}.

The Schr\"odinger equation (\ref{Se}) with the effective
potential (\ref{chi2}) in the dimensionless notation reads
\begin{eqnarray} 
&-&{1\over 2\kappa^2}\tilde\nabla^2\omega+
{(\tilde\nabla\omega)^2\over 4\kappa^2\omega}+\omega Q^2
\nonumber\\
&&=\omega-{t^2\over 1-t^2}\left(
{1\over\sqrt{1-(1-t^4)\omega}}-1\right)\omega.
\label{eq9}
\end{eqnarray}
The terms on the left hand side result from the kinetic energy,
the terms on the right hand side represent the potential.

The Maxwell equation (\ref{Me}) with the current (\ref{Mej})
reads
\begin{equation} 
  -\tilde\nabla^2{\bf Q}_b=-\omega{\bf Q}_A-\omega{\bf Q}_b.
\label{eq10}
\end{equation}
The full quantum velocity is a sum of two terms, ${\bf Q}=
{\bf Q}_A+{\bf Q}_b$, where ${\bf Q}_A$ is any model
velocity field which covers the singular contributions at
vortices,
\begin{equation} 
\tilde\nabla\times{\bf Q}_A=\bar{\bf b}-\Phi_0\sum_{\bf R}
\delta(\tilde{\bf r}-{\bf R}).
\label{singcont}
\end{equation}
In our choice of ${\bf Q}_A$ appears the mean value of the
magnetic field in the superconductor,
\begin{equation} 
\bar{\bf b}=\langle{\bf b}\rangle=
{1\over\Omega}\int d\tilde{\bf r}\,
{\bf b},
\label{meanb}
\end{equation}
therefore, for an ideally periodic vortex lattice ${\bf Q}_A$
is given by the Abrikosov $B_{c2}$ solution.\cite{B97}
The sum over the 2D $\delta$-functions represents the
contributions of the nodes of the wave function in the vortex
centers ${\bf R}$ to the quantum velocity. In this choice
one has $\langle\nabla\times{\bf Q}_b\rangle=0$ so that
$\nabla \times{\bf Q}_b = {\bf b} - \bar{\bf b}$ describes the
spatial modulation of the magnetic field due to the
diamagnetic currents.

\subsection{Fourier representation} 
For the periodic lattice of vortices, it is advantageous to
express all functions by Fourier series,
\begin{eqnarray} 
\omega(\tilde{\bf r})&=&\sum_{{\bf K}\ne 0}a_{\bf K}
(1-\cos{\bf K}\tilde{\bf r}),
\label{Fourexpw}\\
b(\tilde{\bf r})&=&\bar b+\sum_{{\bf K}\ne 0}b_{\bf K}
\cos{\bf K}\tilde{\bf r}.
\label{FourexpQ}
\end{eqnarray}
We choose the direction of vortices along the $z$-axis.
The function $b(\tilde{\bf r})$ is the $z$ component of the magnetic
field ${\bf b}$. Since the system is translationally invariant
along $z$, the vectors $\tilde{\bf r} = (x,y)$ and ${\bf K}$
(reciprocal lattice vectors) are two dimensional.

The special choice, $\omega\propto (1-\cos{\bf K}\tilde{\bf r})$, 
enforces nodes of the wave function, $\omega({\bf R})=0$, at the
positions of the vortex centers, ${\bf R}=(ix_1+jx_2,jy_2)$
with $i,j=0,\pm 1,\pm 2,\ldots$.
For the triangular lattice, two of the nearest neighbor
vortices are at ${\bf R}=(x_1,0)$ and ${\bf R}=(x_2,y_2)$,
where $x_1=2x_2$ and $y_2=\sqrt{3}\,x_2$. The distance
between vortices, $x_1$, is determined by the condition that
each vortex contributes to the mean magnetic field by one
elementary quantum of flux,
$S\bar b=x_1y_2\bar b=\Phi_0=2\pi/\kappa$.
The sums in the Fourier representation run over nonzero
discrete momenta, ${\bf K}=(2\pi/S)(iy_2,jx_1+ix_2)$. In
this choice of the Fourier expansion the mean value of the
amplitude of the wave function reads
\begin{equation} 
\bar\omega=\langle\omega\rangle=\sum_{{\bf K}\ne 0}a_{\bf K}.
\label{meanom}
\end{equation}

  Since ${\bf Q}_b={\bf Q -Q}_A$ is periodic one may write
\begin{equation} 
{\bf Q}(\tilde{\bf r})={\bf Q}_A(\tilde{\bf r})+\sum_{{\bf K}\ne 0}
b_{\bf K}{\hat{\bf z}\times{\bf K}\over K^2}\sin{\bf K}\tilde{\bf r}
\label{eq10F}
\end{equation}
with  $\hat{\bf z}\times{\bf K}\equiv(K_y,-K_x)$ and
$\hat{\bf z}$ is the unit vector along the axis $z$.

\subsection{Simple iteration scheme} 
Now we are ready to specify the iteration scheme for the
Fourier components of the wave function and the
quantum velocity.

The Fourier representation of Eq.~(\ref{eq9}) reads
\begin{equation} 
a_{\bf K}:={4\kappa^2\left\langle(s-2\omega+\omega Q^2+g)
\cos{{\bf K}\tilde{\bf r}}\right\rangle\over K^2+2\kappa^2},
\label{eq11}
\end{equation}
with
\begin{eqnarray} 
s&=&{t^2\over 1-t^2}\left({1\over\sqrt{1-\omega(1-t^4)}}-1\right)
\omega ,
\label{pdef}\\
g&=&{(\tilde\nabla\omega)^2\over 4\kappa^2\omega}.
\label{gdef}
\end{eqnarray}

The Fourier representation of Eq.~(\ref{eq10}) reads
\begin{equation} 
b_{\bf K}:=-{2\left\langle(\omega b+\bar\omega(b-\bar b)+p)
\cos{{\bf K}\tilde{\bf r}}\right\rangle\over K^2+\bar\omega},
\label{eq12}
\end{equation}
with
\begin{equation} 
p=(\nabla\omega\times{\bf Q})\hat{\bf z}=Q_x{\partial\omega
\over\partial y}-Q_y{\partial\omega\over\partial x}.
\end{equation}

Within a simple iteration scheme for given values of $t$,
$\bar b$, and $\kappa$, one starts from the Abrikosov $B_{c2}$
solution or some other values of $a_{\bf K}$ and $b_{\bf K}$.
In the step ({\bf a}) one evaluates ${\bf a}_{\bf K}$ from
(\ref{eq11}) and upgrades  $\omega(x,y)$ according to
(\ref{Fourexpw}). In the step ({\bf b}) one evaluates 
$b_{\bf K}$ from (\ref{eq12}) and upgrades $b(x,y)$ and 
$Q(x,y)$ from (\ref{FourexpQ}) and (\ref{eq10F}). 
The iteration scheme ({\bf a}), ({\bf b}), ({\bf a}), 
({\bf b}), $\ldots$ then leads to the periodic solution 
of Eqs.~(\ref{eq9}) and (\ref{eq10}).

\subsection{Accelerated iteration scheme} 

As shown in Ref.~\onlinecite{B97}, the convergence is
accelerated if the amplitude of the wave function is
optimized after each use of equation (\ref{eq11}). Here we
show how to make this optimization within the Bardeen set
of equations.

Assume a change of the wave function $\omega(x,y)$ which
maintains its shape but modifies its amplitude,
\begin{equation} 
\omega(x,y) := (1+c)\omega(x,y),
\label{resc}
\end{equation}
i.e., the old value $\omega$ (the right hand side) obtained from
(\ref{eq11}) is replaced by a new value.

The constant $c$ has to be found at each iteration step from
the minimum of the free energy. Since we neglect the Coulomb
interaction in this part of the treatment, we can also eliminate
the internal energy $U$. The free energy normalized as
\begin{equation} 
\tilde f={f_s-U+f_{\rm kin}+f_M\over{1\over 4}\gamma
T_c^2(1-t^2)(1-t^4)},
\label{fe2}
\end{equation}
in the dimensionless representation reads
\begin{eqnarray} 
\tilde f&=&-{\omega\over 1-t^2}-{2 t^2\sqrt{1-\omega(1-t^4)}
\over(1-t^2)(1-t^4)}
\nonumber\\
&&+(\tilde\nabla\times{\bf Q}-b_a)^2+\omega Q^2+
{(\tilde\nabla\omega)^2\over 4\kappa^2\omega}.
\label{fe1}
\end{eqnarray}
When we substitute (\ref{resc}) into (\ref{fe1}), the
variation with respect to $c$, $\partial\langle\tilde f
\rangle/\partial c=0$, yields the condition for $c$ as
\begin{eqnarray} 
&&\left\langle-{\omega\over 1-t^2}+\omega Q^2+
{(\tilde\nabla\omega)^2\over 4\kappa^2\omega}
\right\rangle
\nonumber\\
&&=-\left\langle{t^2\over 1-t^2}{\omega\over
\sqrt{1-(1+c)\omega(1-t^4)}}\right\rangle .
\label{rescimpl}
\end{eqnarray}

Condition (\ref{rescimpl}) is not convenient for the numerical
treatment. When the starting value of the wave function is
reasonable, or after a few iteration steps
({\bf a}), ({\bf b}), ({\bf a}), ({\bf b}),
the correction $c$ will be small, $c\ll 1$,
so that the linear approximation of (\ref{rescimpl}) is
sufficient,
\begin{equation} 
c=-{2\left\langle s-\omega+\omega Q^2+g\right\rangle\over
t^2(1+t^2)\left\langle\omega^2
\left(1-\omega(1-t^4)\right)^{-{3\over 2}}\right\rangle}.
\label{rescimpllin}
\end{equation}

  As a third iteration  step ({\bf c}) we thus may use
Eq.~(\ref{resc}) with $c$ given by (\ref{rescimpllin}).
The iteration procedure we use starts from a preliminary
adjustment of the wave function by a few steps,
({\bf a}), ({\bf c}), ({\bf a}), ({\bf c}),
$\ldots$, putting $b_{\bf K} \equiv 0$. After that the
full iteration scheme ({\bf a}), ({\bf c}), ({\bf b}),
({\bf a}), ({\bf c}), ({\bf b}), $\ldots$ is applied
yielding all Fourier coefficients $a_{\bf K}$, $b_{\bf K}$.

\subsection{Magnetic properties} 

In Figs.~1-4 we present some numerical results to
illustrate the properties of the Bardeen equations. As one
can see from the dimensionless equations (\ref{eq9}) and
(\ref{eq10}), the behavior of the system is determined by
a single material parameter, the GL parameter $\kappa$.
We assume Niobium doped with non-magnetic impurities of
a density giving the GL parameter $\kappa_0=1.5$.\\

  \begin{figure}[h]   
  \centerline{\parbox[c]{8cm}{
  \psfig{figure=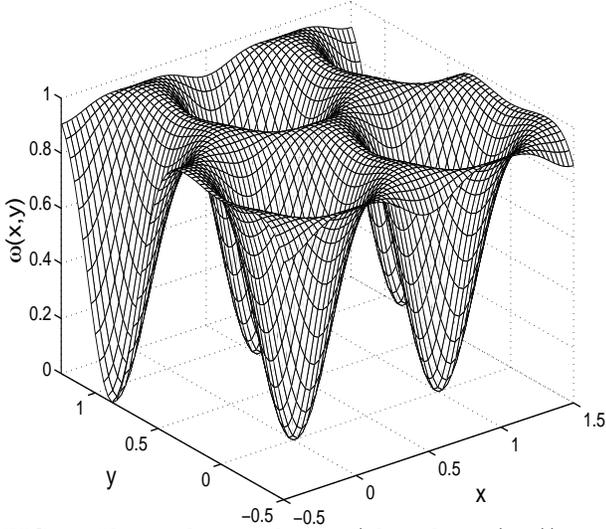,width=8cm,height=7cm}}}
\caption{The condensate density (plotted as $\omega(x,y)$)
in the triangular lattice for temperature $t=0.5$,
magnetic induction $\bar B/B_{c2} =0.5$, and GL parameter
$\kappa_0=1.5$. In the vortex centers the condensate density
$n_s(x,y)/n = (1-t^4) \omega(x,y)$ goes to zero. Between the
vortices $\omega(x,y)$ approaches its equilibrium value 1
(which would be constant in the absence of a magnetic field)
yielding $n_s^{\rm eq}/n = 1-t^4 = 0.94$.}
\label{p1}
\end{figure}

Figure~\ref{p1} shows a fishnet plot of the condensate density
$n_s(x,y)/n= (1-t^4)\omega(x,y)$ for temperature $t=T/T_c=0.5$ and
the mean magnetic field $\bar B= 0.5\,B_{c2}$. The dips in $n_s$
correspond to the nodes of the wave function $\psi$ located at the
vortex centers. The condensate density reaches its maximum between
the vortices.\\

\begin{figure}  
  \centerline{\parbox[c]{8cm}{
  \psfig{figure=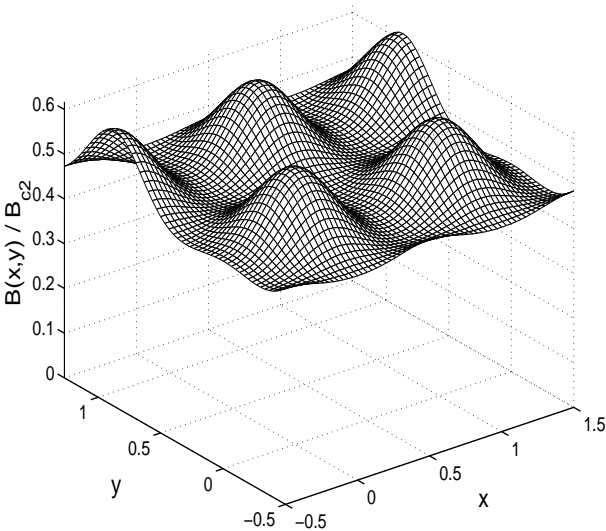,width=8cm,height=7cm}}}
\vskip 4mm
\caption{The magnetic field in units of the upper critical field
$B_{c2}$ for $t=0.5$, $\bar B/B_{c2}=0.5$, and $\kappa_0=1.5$ as
in Fig.~\ref{p1}. $B(x,y)$ reaches its maximum $B_{\rm max}$ at
the vortex centers.}
\label{p2}
\end{figure}

Note that the condensate density is smaller than its
non-magnetic value, $n_s^0=n(1-T^4/T_c^4)$, also on the borders
of the elementary cells where the current is zero. This shows
that non-local effects given by gradient corrections, e.g.
the second term of (\ref{B3}), are important not only at
the vortex core but also between the vortices.

A complementary picture offers the plot of the magnetic
field $B$ presented in Fig.~\ref{p2}. The magnetic field
reaches its maximum value, $B_{\rm max}$, at the vortex
centers. This
maximum field is very close to but slightly higher than
the applied field $B_a$ because the superconductor tries to
expel the magnetic field and compresses it into vortices.
The magnetic pressure on the condensate is one of the forces
balanced by the electric field.

\begin{figure}  
  \centerline{\parbox[c]{8cm}{
  \psfig{figure=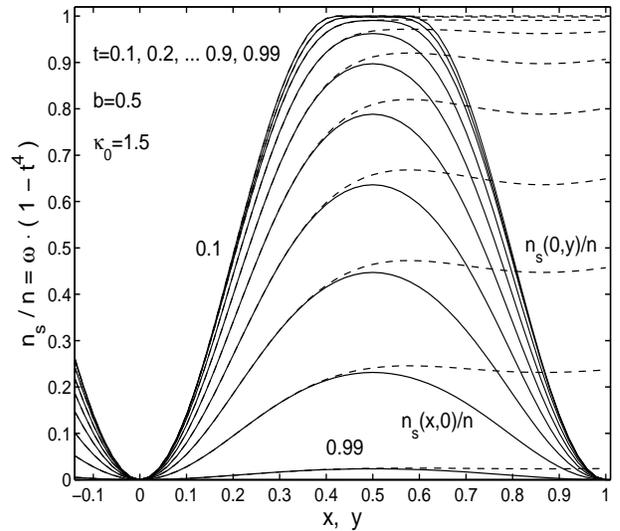,width=8cm,height=7cm}}}
\caption{Profiles of the condensate density $n(x,y)$ at various
temperatures for $\bar B/B_{c2}=0.5$ and $\kappa_0 = 1.5$.
The solid lines show cuts along the $x$-axis, and the dashed
lines along the $y$-axis.}
\label{p3}
\end{figure}
The temperature dependence of the condensate density
$n_s(x,y)/n = (1-t^4) \omega(x,y) $ is shown in Fig.~\ref{p3}.
As one expects, the density of the condensate decreases as
the temperature approaches the critical value, $t\to 1$.
The dominant part of this decrease can be attributed
to the reduced fraction of the condensate expressed by the
factor $(1-t^4)$. We have numerically checked that near the
critical temperature the condensate density results identical
to the solution of the standard GL theory.

\begin{figure}   
  \centerline{\parbox[c]{8cm}{
  \psfig{figure=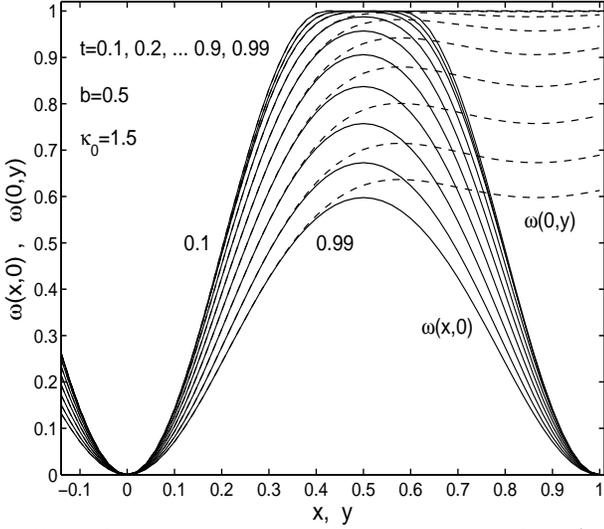,width=8cm,height=7cm}}}
\caption{Profiles of the reduced condensate density $\omega(x,y)$
at various temperatures $t=T/T_c$ for $\bar B/B_{c2}=0.5$ and
$\kappa_0 = 1.5$ as in Fig.~\ref{p3}.}
\label{p4}
\end{figure}

In Fig.~\ref{p4}, the density of the condensate is normalized
to its value in absence of the magnetic field,
$n_s(x,y)/n_s^0=\omega(x,y)$. The suppression of $\omega(x,y)$
in the region between the vortices is completely due to the magnetic
field. One can see that at lower temperatures the condensate
is less suppressed than it would result from the GL theory, where
the latter one is equal to the curve at $t=0.99$.
This follows from the fact that the condensation energy
${\cal F}_s$ increases with the condensate density slower
than the quadratic function of the GL theory.

\subsection{Virial theorem} 

In the above treatment, the magnetic field was specified
by the mean value $\bar B$ of the magnetic induction in the
sample. Macroscopic magnetic properties of the system,
however, are given by the magnetization $M$. Let us link
these two quantities.

For simplicity we assume that the sample is an infinite
cylinder in the direction of the applied field. In this
longitudinal geometry one has
$\bar B=B_a+M$. Since the non-local terms of the free
energy are terminated at the second-order derivatives (the
term called kinetic energy), we can conveniently use
the virial theorem derived by Doria, Gubernatis and
Rainer\cite{DGR89} and generalized by Klein and
P\"ottinger\cite{KP91}, to evaluate the applied magnetic
field $B_a$.

The idea of the virial theorem is as follows. Let us introduce
a parameter $\iota$ which scales coordinates $x$ and $y$.
With this scaling one can generate a new wave function
$\omega'({\bf r})=\omega(\iota{\bf r})$. Since the mean magnetic
field is given by the density of vortices, it scales as
$\bar B'({\bf r})=\iota^{-2}\bar B(\iota{\bf r})$. We rescale
all magnetic fields with $\iota^{-2}$ except for the applied
field $B_a$ which is an external parameter.

  From ${\bf B}=\nabla\times{\bf A}$ one can see that the vector
potential scales as ${\bf A}'({\bf r})=\iota^{-1}{\bf A}(\iota
{\bf r})$, i.e.,  in the same way as a gradient. Accordingly,
the density of kinetic energy scales with $\iota^{-2}$. The
mean value of free energy corresponding to the new
wave function reads
\begin{eqnarray} 
\langle f'\rangle&=&\left\langle -{\omega\over 1-t^2}-
{t^2\sqrt{1-\omega(1-t^4)}\over(1-t^2)(1-t^4)}\right\rangle
\nonumber\\
&+&
\iota^{-2}\left\langle\omega Q^2+{(\tilde\nabla\omega)^2\over
4\kappa\omega}\right\rangle +
\left\langle(\iota^{-2}\tilde\nabla\times{\bf Q}-b_a)^2
\right\rangle .
\nonumber\\
\label{fe3}
\end{eqnarray}
The condensation energy (the first term) is independent
of the scaling. The kinetic energy (the second term)
scales with $\iota^{-2}$. The magnetic energy (the third
term) has three contributions: $\langle b^2\rangle=
\langle(\nabla\times{\bf Q})^2\rangle$ which scales with
$\iota^{-4}$; $-2\bar bb_a=-2\langle(\nabla\times{\bf Q})
b_a\rangle$ which scales  with $\iota^{-2}$; and $b_a^2$
which is independent of the scaling.

Since the scaling deforms the wave function and the
internal magnetic field from their equilibrium values, the
free energy $\langle f'\rangle$ is greater than the free
energy $\langle\tilde f\rangle$. For $\iota=1$ the free
energy $\langle f'\rangle$ reaches its minimum being equal
to $\langle\tilde f\rangle$. This minimum is given by a
variation with respect to $\iota$,
\begin{equation} 
\left.{\partial\over\partial\iota}\langle f'\rangle
\right|_{\iota=1}=0.
\label{variota}
\end{equation}
Condition (\ref{variota}) in the explicit form,
\begin{equation} 
2\bar bb_a=\left\langle\omega Q^2+{(\tilde\nabla\omega)^2\over
4\kappa\omega}\right\rangle+
2\langle(\nabla\times{\bf Q})^2\rangle ,
\label{bavir}
\end{equation}
is called the virial theorem. Since $\bar b$, $\omega$ and
${\bf Q}$ are known, the virial theorem (\ref{bavir}) provides
us with the value of the applied magnetic field $b_a$ without
having to take the derivative of the computed free energy.

A convenient form of the virial theorem valid only for the
Bardeen equations makes use of the Schr\"odinger
equation (\ref{eq9}) from which follows
\begin{equation} 
\left\langle\omega Q^2+{(\nabla\omega)^2\over 4\kappa\omega}
\right\rangle=\langle\omega-s\rangle 
\label{means}
\end{equation}
with $s$ from (\ref{pdef}). The applied magnetic field then reads
\begin{equation} 
b_a={\langle 2b^2+\omega-s\rangle\over 2\bar b}.
\label{applb}
\end{equation}

\begin{figure}  
  \centerline{\parbox[c]{8.2cm}{
  \psfig{figure=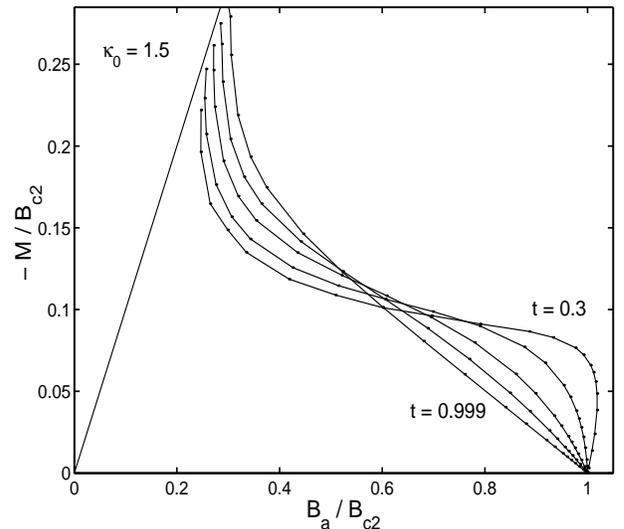,width=8cm,height=7cm}}}
\caption{The magnetization $-M=B_a -\bar B$ as a function of the
applied magnetic field $B_a$ in units of the upper critical field
$B_{c2}$ at temperatures $t =0.999 $, 0.85, 0.7, 0.5, 0.3 for
$\kappa_0 = 1.5$.} \label{p5}
\end{figure}
The magnetization $-M=B_a -\bar B$ as a function of the applied
magnetic field $B_a$ is shown in Fig.~\ref{p5} for different
temperatures. At temperatures close to $T_c$ the magnetization
follows the line well known from the GL theory. Below the lower
critical field $B_{c1}$ the system is in the Meissner state
and $-M=B_a$. Above $B_{c1}$ the magnetization decreases and
linearly vanishes at $B_{c2}$ where the system undergoes
a transition into the normal state.

At very low temperatures, the magnetization is deformed into
an S-shape. The slope of the decrease, $\partial M/\partial B_a$,
close to $B_{c2}$ increases with decreasing temperature and at
a certain temperature $T_a$ becomes infinite. Below $T_a$,
the magnetic behavior of the system achieves an anomal feature.
As the magnetic field is lowered from some high value, the
system undergoes a first order transition from zero to
a finite magnetization at a field which is above $B_{c2}$.
Since the free energy of the system with finite magnetization
is lower than the free energy of the normal state, the
system jumps to a finite magnetization as soon as the applied
magnetic field allows for such solution.

Such anomalous magnetic transition has been observed by Ehrat
and Rinderer\cite{ER68,ER74} for Lead doped with Niobium. In
spite of this experimental result we believe that the first
order transition seen in Fig.~\ref{p5} is an artifact of the
Bardeen approximation. Indeed, detailed theoretical
discussions\cite{FU71,J71a,J71b} of this anomalous behavior
point to the important role of scattering on impurities.
This mechanism is absent in the Bardeen approximation. 

The temperature $T_a$ can be determined from the Bardeen
equations. Close to the critical field $B_{c2}$ the density
of condensate is small and one can expand the effective
potential (\ref{chi2}) into the GL form (\ref{chi3}) with
coefficients
\begin{equation}
\alpha={\gamma\over 2n}(T^2-T_c^2),~~~~~~~~
\beta={\gamma T^2\over 2n^2}.
\label{Bc2beta1}
\end{equation}
For these asymptotic values one can introduce an asymptotic
GL parameter,\cite{W96}
\begin{equation}
\kappa_{\rm as}=\sqrt{m^2\beta\over 2\mu_0\hbar^2 e^2}.
\label{Bc2beta2}
\end{equation}
As one can see from (\ref{Bc2beta1}), this asymptotic GL
parameter decreases with the temperature,
\begin{equation}
\kappa_{\rm as}=\kappa_0 t.
\label{Bc2beta3}
\end{equation}
The transition temperature $T_a$ appears when the effective
GL parameter equals to $1/\sqrt{2}$, i.e.,
\begin{equation}
T_a={T_c\over\sqrt{2}\kappa_0}.
\label{Bc2beta4}
\end{equation}
For $\kappa_0=1.5$ one finds $T_a=0.47\,T_c$. We expect that
one should be cautious about results of the Bardeen equations
below $T_a$.

\begin{figure}  
  \centerline{\parbox[c]{8cm}{
  \psfig{figure=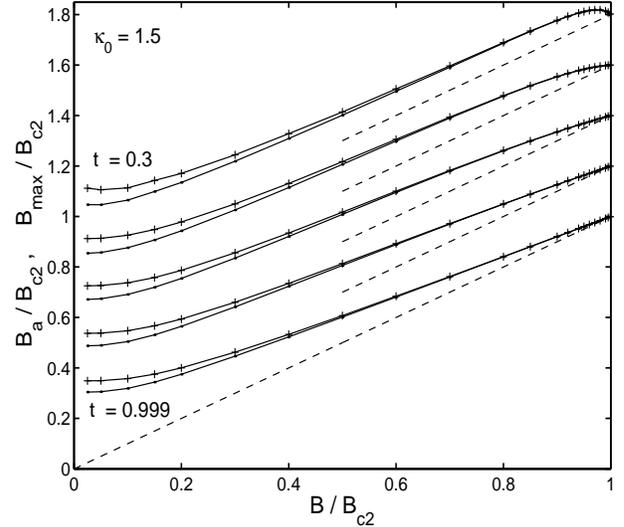,width=8cm,height=7cm}}}
\vskip 2mm \caption{The applied field $B_a$ (solid lines with
dots) and the field $B_{\rm max}$ in the vortex center (solid
lines with crosses) plotted versus the induction $\bar B$ for
$\kappa_0 = 1.5$ at temperatures $t=0.999$, 0.85, 0.7, 0.5, 0.3 as
in Fig.~\ref{p5}. For clarity, each line pair of the next
temperature is shifted up by 0.2. The dashed lines indicate the
large-$\kappa$-limit $B_a= \bar B$.} \label{p6}
\end{figure}

Let us return to features related to the electrostatic forces.
As mentioned above, in the vortex core the magnetic field $B(x,y)$
is compressed and thus exceeds the value of the applied field
$B_a$. In Fig.~\ref{p6} we compare the applied field with the field
$B_{\rm max}$ in the center of the vortex. For all
temperatures the compression is stronger at lower magnetic fields.

\section{Electrostatic potential and charge in
the Abrikosov lattice} \label{VI}

The electrostatic potential, $\varphi$, together with the
Sorokin thermodynamic potential, $w_s$, control the motion
of Cooper pairs. Indeed, the total effective potential
acting on the Cooper pairs is $\chi=e^*\varphi+2w_s$.
The separation of the effective potential $\chi$ into its
electrostatic and thermodynamic components sheds a
light on the role of the electrostatic potential in the
Schr\"odinger equation (\ref{Se1}) or (\ref{Se}).

\subsection{Electrostatic potential} 

The electrostatic potential in the vortex lattice is given
by the screened Poisson equation (\ref{chi6}). For simplicity
we neglect the screening, putting
$\lambda^2_{TF}\nabla^2e\varphi = 0$.
This approximation is justified below.

To be compatible with the above notation, we define a
dimensionless electrostatic potential,
\begin{equation} 
\phi={en\over{1\over 4}\gamma T_c^2(1-t^2)(1-t^4)}
\varphi .
\label{nodimphi}
\end{equation}
With the screening neglected one finds from (\ref{chi6})
\begin{equation} 
\phi=s-\omega+C_1\omega+C_2\sqrt{1-(1-t^4)\omega},
\label{nscPeF}
\end{equation}
with temperature dependent factors
\begin{equation} 
C_1={1\over 1-t^2}
{\partial\ln\varepsilon_{\rm con}\over\partial\ln n},
\label{PeFour1}
\end{equation}
and
\begin{equation} 
C_2={1\over 1-t^2}{\partial\ln\gamma\over\partial\ln n}.
\label{PeFour2}
\end{equation}
The term $s-\omega$ in (\ref{nscPeF}) corresponds to $\chi
|\psi|^2/n$ in (\ref{chi6}). The $C_{1,2}$ terms
correspond to the second and third terms of (\ref{chi6}),
respectively.\\

\begin{figure}  
  \centerline{\parbox[c]{8.0cm}{
  \psfig{figure=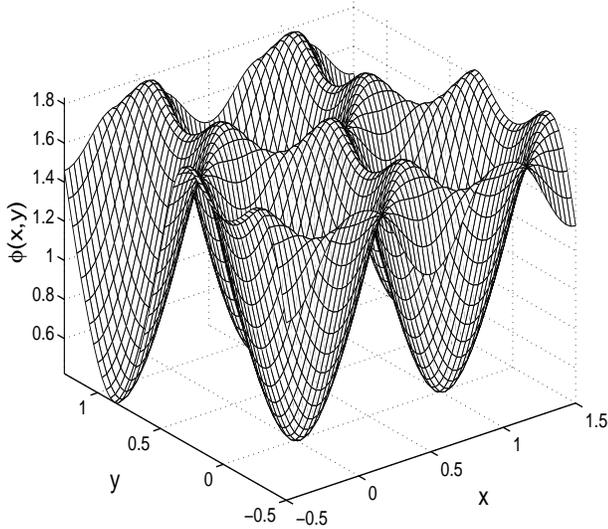,width=8cm,height=7cm}}}
\vskip 4mm
\caption{The electrostatic potential $\phi(x,y)$, Eq.~(101).
The temperature $t=0.5$, the magnetic field $\bar B/B_{c2} =0.5$
and the GL parameter $\kappa_0 =1.5$ are identical to the values
used in Figs.~\ref{p1} and \ref{p2}. The thermodynamic
coefficients $C_1$ and $C_2$ are specified in Tab.~1.}
\label{p7}
\end{figure}
Figure~\ref{p7} shows a fishnet plot of the electrostatic
potential. The potential reaches its minimum at the vortex
centers, i.e., it attracts electrons to vortices.

\begin{figure}  
  \centerline{\parbox[c]{8cm}{
  \psfig{figure=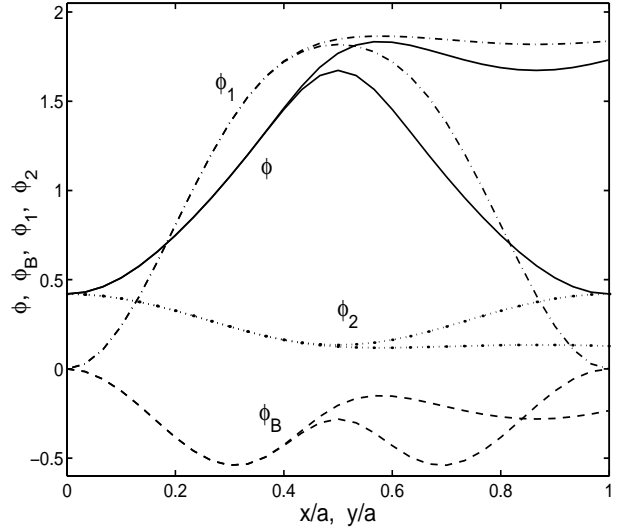,width=8cm,height=7cm}}}
\vskip 2mm
\caption{The components of the electrostatic potential in the
vortex lattice of spacing $a$ according to
Eq.~(\protect\ref{nscPeF}). The individual potentials are: the
total potential $\phi$ (solid lines), the Bernoulli potential
$\phi_B =s-\omega$ (dashed lines), the condensation potential
$\phi_1 =C_1\omega$ (dashed-dotted lines), and the normal
thermodynamic potential $\phi_2 =C_2\sqrt{1-(1-t^4)\omega}$
(dotted lines). The splitting of lines at larger distances
characterizes the $x$-direction (lower curves) or $y$-direction
(upper curves). Parameters as in Fig.~7.}
\label{p8}
\end{figure}

The total electrostatic potential $\phi$ is composed of
three components: the Bernoulli potential $\phi_B=s-\omega$,
the contribution due to the condensation energy $\phi_1=
C_1\omega$ and the reduced normal state thermoelectric
potential $\phi_2=C_2\sqrt{1-(1-t^4)\omega}$. Individual
components are compared in Fig.~\ref{p8}.

The Bernoulli potential $\phi_B$ shown in Fig.~\ref{p8} is
negative.  Due to the quasiparticle screening, the Bernoulli
potential reaches zero at the center of the vortex. With
respect to the center of the vortex, the forces corresponding
to the Bernoulli potential are repulsive inside the core
while they are attractive outside.

The potential $\phi_1$ caused by the density dependence of the
condensation energy is positive. Being proportional to the
density of condensate, it has a minimum at the vortex center
where it reaches zero. For Niobium this contribution is
dominant since the coefficient $C_1=1.9$ is rather large
compared to the coefficients of other contributions.

The normal state thermodynamic potential $\phi_2$ is also
positive giving the only non-zero contribution at the vortex
center. One can see that $\phi_2$ reduces the total potential
since it has the maximum at the vortex core and falls outside.
Its coefficient $C_2=0.42$ is about four times smaller than
$C_1$, therefore this term cannot cancel the potential
$\phi_1$.

We want to stress that even at temperature $t=0.5$ when
96\% of electrons are in the condensate, the thermodynamic
correction to the electrostatic potential cannot be
neglected. This result contradicts the temperature 
dependence of the thermodynamic correction derived by 
Rickayzen\cite{R69}, see Eq.~(\ref{Ric4}). Within the
hydrodynamic picture one can show that the limit of weak
currents adopted by Rickayzen is responsible for this 
disagreement. In this limit, the effect of the diamagnetic
current on the condensate density $n_s$ vanishes as 
$T\to 0$, see Eq.~(\ref{nsprime}). The temperature 
dependence of the thermodynamic correction merely
reflects the temperature dependence of $n_s'$. The limit
of weak current does not apply to the vortex core. In the 
center of the vortex core, the condensate density has to 
go to zero keeping the magnitude of the thermodynamic 
correction appreciable at any temperature. 

\subsection{Effective potential} 

\begin{figure}  
  \centerline{\parbox[c]{8cm}{
  \psfig{figure=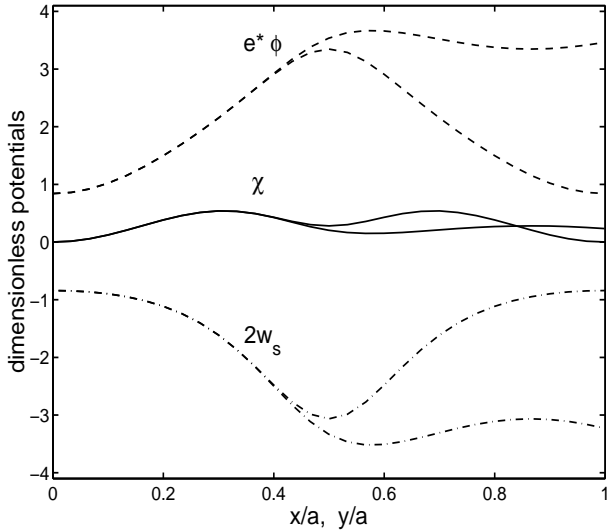,width=8cm,height=7cm}}}
\vskip 2mm
\caption{The effective potential $\chi=e^*\varphi+2w_s$ 
(full lines),
the electrostatic potential acting on the Cooper pair
$e^*\varphi$ (dashed lines) and the thermodynamic potential
$2w_s$ (dash-dotted lines) for $\kappa_0 = 1.5$. Parameters
and presentation as in Fig.~\protect\ref{p8}.}
\label{p9}
\end{figure}

To enlighten the role of the electrostatic potential in the
balance of forces in superconductors, we compare the effective 
potential $\chi$, the electrostatic potential acting on the 
Cooper pair $e^*\varphi$, and the thermodynamic potential of 
Sorokin $2w_s$ in Fig.~\ref{p9}. All these contributions are 
in dimensionless units corresponding to (\ref{nodimphi}).

One can see that the electrostatic potential is not a small
correction to the effective potential of a thermodynamic
origin. The amplitude of the electrostatic potential is
about an order of magnitude larger than the amplitude of
the effective potential $\chi$. Accordingly, the effective
potential $\chi=e^*\varphi+2w_s$ results from a strong 
compensation of the thermodynamic potential $2w_s$ and the 
electrostatic potential $e^*\varphi$.

\subsection{Charge} 

The distribution of the charge in the vortex lattice is given
by the Poisson equation, $\rho=-\epsilon\nabla^2\varphi$.
We introduce a dimensionless charge,
\begin{equation} 
\tilde\rho={\rho\over en} ,
\label{char1}
\end{equation}
which measures the relative deviation of the charge density
from the crystal value. In dimensionless representation
the Poisson
equation reads
\begin{equation} 
\tilde\rho=-C_3{\lambda_{TF}^2\over\lambda_{\rm Lon}^2}
\tilde\nabla^2\phi,
\label{char2}
\end{equation}
with
\begin{equation} 
C_3={2{\cal D}\varepsilon_{\rm con}\over n^2}
(1-t^2)(1-t^4).
\label{char3}
\end{equation}

\begin{figure}   
  \centerline{\parbox[c]{8cm}{
  \psfig{figure=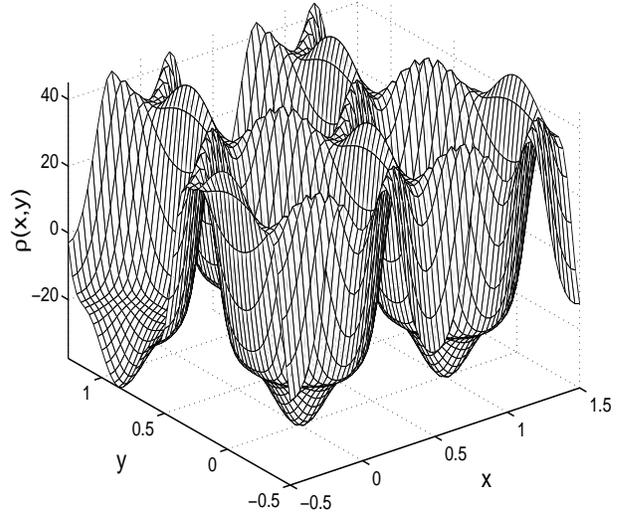,width=8cm,height=7cm}}}
\vskip 4mm
\caption{The function $-\nabla^2\phi$ proportional to the charge
density $\rho(x,y)$. The amplitude of the
dimensionless charge density is $\tilde\rho =9.5\,10^{-11} \cdot
(1-t^2) (1-t^4)^2\tilde\nabla^2\phi$. Same parameters as
in Figs.~7, 8, 9.}
\label{p10}
\end{figure}
Figure~\ref{p10} shows a fishnet plot of the charge
distribution. In the vortex core the charge is depleted,
the missing charge is distributed between vortices.

A striking feature is the very rapid change of the charge
sign at the distance about 0.4 from the vortex center. While
the charge in the core is rather flat, its spatial variation
between vortices is quite strong. This picture of the
charge distribution is just opposite to the one assumed
by Kumagai, Nozaki and Matsuda\cite{KNM} who
expected a flat charge distribution between vortices.
Comparing these two pictures, however, one has to keep
in mind that Kumagai {\em et al} discuss YBCO with
$\kappa\sim 100$ while Fig.~\ref{p10} presents the case
of $\kappa_0=1.5$ in Niobium.

\begin{figure}   
  \centerline{\parbox[c]{8cm}{
  \psfig{figure=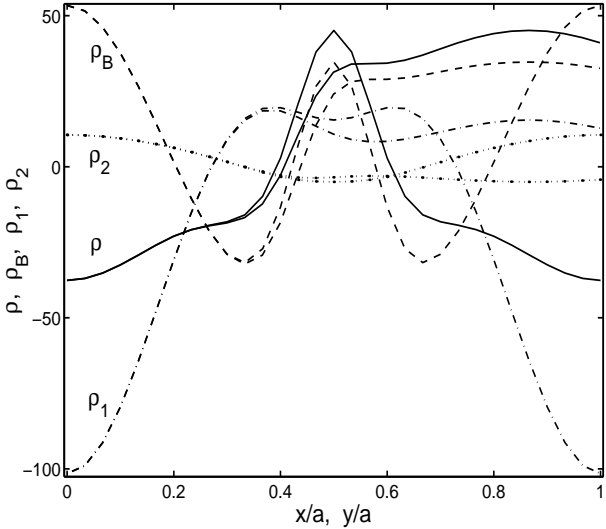,width=8cm,height=7cm}}}
\vskip 2mm
\caption{The components of the charge density in the notation
of Fig.~(\protect\ref{p8}) for same parameters. The individual
charge densities are:
the total $\rho$ (solid lines), the Bernoulli $\rho_B$ (dashed
lines), the condensation part $\rho_1$ (dash dotted lines),
and the normal thermodynamic part $\rho_2$ (dotted lines).}
\label{p11}
\end{figure}

The particular shape of the charge seen in Fig.~\ref{p10}
results from the interplay between the Bernoulli potential
$\phi_B$ and the potential $\phi_1$ due to the condensation
energy. In Fig.~\ref{p11}, we show the charge density
decomposed into contributions corresponding to individual
potentials, $\rho_i\propto\nabla^2\phi_i$.

The charge distribution $\rho_B$ corresponding to the
Bernoulli potential $\phi_B$ has its maximum at the center
of the vortex. In Ref.~\onlinecite{KLB01}, where only the
non-local (quantum) Bernoulli potential has been assumed,
there is a minimum of the charge density in the center of
vortex. The local maximum seen in Fig.~\ref{p11} follows
from the quasiparticle screening not assumed in
Ref.~\onlinecite{KLB01}.

The amplitudes of the contributions $\rho_1$ and $\rho_2$
depend on the constants $C_1$ and $C_2$, which strongly
depend on the material in question. For Niobium one has
$C_1= 1.9$ and $C=0.42$; therefore $\rho_1$ dominates. In
Appendix~A one can see that both constants $C_1$ and
$C_2$ are proportional to the slope of the density of
states at the Fermi level. In general one can say that the
amplitude of $\rho_2$ is smaller than the amplitude of
$\rho_1$ and the two contributions have opposite signs
of the periodic parts. We note that due to the large
value of $C_1$ and small $C_2$, the total charge has
a minimum in the center of the vortex.

\subsection{Screening and the quasi-neutral approximation} 

  For Niobium, the Thomas-Fermi screening length is
very small,
\begin{equation} 
{\lambda_{TF}^2\over\lambda_{\rm Lon}^2}=2.5\, 10^{-6}(1-t^4).
\label{char4}
\end{equation}
One can thus neglect the screening,
$\lambda_{TF}^2\lambda_{\rm Lon}^{-2}\tilde\nabla^2
\phi\ll\phi$. Indeed, the Laplace operator in the Fourier
representation is $\tilde\nabla^2\to K^2=(4\pi/\sqrt{3})
\kappa \bar b(i^2+ij+j^2)$. Since $\kappa$ is of the order
of unity, $\bar b<1$ and the number of needed Fourier
components is also limited, $i,j<100$, the screening is
negligible for all Fourier components considered.

  For Niobium, the factor
\begin{equation} 
{2{\cal D}\varepsilon_{\rm con}\over n^2}=3.8\,10^{-5}
\label{char5}
\end{equation}
which determines $C_3$ in (\ref{char2}), is also very small. 
Similarly small value can be expected for any conventional 
superconductor.
It leads to relative charges of the order of $10^{-10}$.
The quasi-neutral approximation, $\gamma(n)\approx
\gamma(n_0)$ etc., is thus well justified when one
solves for the wave function and the vector potential.

\section{Conclusions}  \label{VII}

In this paper we have discussed the electrostatic potential
in the Abrikosov lattice of vortices. To this end we have
derived a set of three Ginzburg-Landau equations which
include the Maxwell equation for the vector potential, the
Schr\"odinger equation for the wave function, and the
Poisson equation for the electrostatic potential. These
equations determine the minimum of the free energy made of
four components: the condensation energy of Gorter and
Casmir; the quantum kinetic energy of Ginzburg and Landau;
the magnetic free energy of Helmholtz; and the Coulomb
energy.

The marriage of the Gorter-Casimir two-fluid model with the
Ginzburg-Landau theory has been suggested earlier by
Bardeen who has also discussed properties of this theory
at different temperature limits. We have employed his
approach as it offers a very simple extension of the GL theory
towards low temperatures. As our results document, this
extended theory can be treated with standard numerical tools
of the GL theory.

With the electrostatic interaction included, the effective
potential acting on the superconducting condensate is
naturally a sum of the electrostatic potential and the
thermodynamic potential. One can say that the electrostatic
potential (over-)screens the thermodynamic potential
leaving a relatively small effective potential.

In spite of the very important role of the electrostatic potential
among forces acting on the condensate, the electrostatic
potential can be eliminated from the Ginzburg-Landau theory
so that one has to solve a set of two, not three, equations.
This simplification is possible by two reasons. First, the
charge modulation which corresponds to this potential, is
so small on the scale of the charge density in metals that
one can neglect its effect on local values of material
parameters.

The second reason is more fundamental. As noticed by
van~Vijfeijken and Staas, there is a force between the
condensate and the normal electrons. This force keeps the
normal electrons at rest, i.e., it balances the electric
field having an equal amplitude and the opposite orientation.
The force of van~Vijfeijken and Staas is an exclusive function
of the condensate density. Accordingly, one can express the
electric force or the electrostatic potential as a function of
the condensate density. In this way the electrostatic potential
can be unified with the thermodynamic potential into an
effective potential of GL-type.

In the numerical treatment we have used the parameters of
Niobium. Our choice of this conventional material was
determined by known empirical rules needed to predict
amplitudes of the individual contributions to the electrostatic
potential. We expect that other d-band superconductors
behave similarly.

Finally, we would like to stress that the presented theory
is simplified in many directions. First, it is restricted to
isotropic materials. We have omitted all features of the band
structure except for the density of states and its slope on
the Fermi level. Second, the two-fluid model of Gorter and
Casimir describes only gross features of the thermodynamics
of superconductors. Third, the gradient approximation of the
Ginzburg-Landau theory is justified only close to the critical
temperature, at low temperatures one has to take the
kinetic energy of Ginzburg and Landau as an ad hoc
approximation. In the future, we plan to address layered
structures and use a more general form of the
Gorter-Casimir model.

\appendix
\section{Estimate of material parameters for Niobium} 
In this appendix we estimate material parameters, $\partial
\gamma/\partial n$ and $\partial\varepsilon_{\rm con}/\partial
n$, which determine the electrostatic potential in the
superconductor, see (\ref{chi6}) . To be specific we assume
Niobium.

\subsection{Coefficient $\partial\gamma/\partial n$} 
The linear coefficient of the specific heat $\gamma$ is linked
to the density of states $\cal D$ per spin and unitary volume,
\begin{equation} 
\gamma={2\over 3}\pi^2k_B^2{\cal D}.
\label{M1}
\end{equation}
It is advantageous to express the density derivative of $\gamma$
in terms of the energy derivative of the density of states. Using
$\partial E_F/\partial n=1/2{\cal D}$ we find
\begin{equation} 
{\partial\gamma\over\partial n}={1\over 3}\pi^2k_B^2
{\partial\ln{\cal D}\over\partial E_F}.
\label{M2}
\end{equation}

The density of states $\cal D$ includes the mass renormalization
due to the electron-phonon interaction,\cite{KSK74}
\begin{equation} 
{\cal D}={\cal D}_0(1+\lambda),
\label{M3}
\end{equation}
where ${\cal D}_0$ is a bare density of states and
$\lambda$ is the coupling parameter. The value and
the energy derivative of ${\cal D}_0$ is provided by
{\em ab initio} studies of Niobium.\cite{EK77}

The value of the coupling parameter $\lambda$ is found
comparing $\cal D$ from the experimental $\gamma$ with
the theoretical ${\cal D}_0$. The energy derivative of
$\lambda$, however, is not provided in the literature. To
estimate the derivative of $\lambda$ we write it as a product,
\begin{equation} 
\lambda={\cal D}_0V,
\label{M4}
\end{equation}
where $V$ is the BCS interaction.

According to trends found from the effects of impurities on the
critical temperature and the specific heat, the major changes
of $\lambda$ follow from the density of states while the BCS
interaction $V$ remains nearly constant.\cite{VD76} As a
first approximation we thus assume
\begin{equation} 
{\partial V\over\partial n}=0~~~~~{\rm or}~~~~~
{\partial V\over\partial E_F}=0.
\label{M5}
\end{equation}

Now we can complete the estimate of
$\partial\gamma/\partial n$. From (\ref{M3}-\ref{M5}) follows
\begin{equation} 
{\partial{\cal D}\over\partial E_F}=(1+2\lambda)
{\partial{\cal D}_0\over\partial E_F},
\label{M6}
\end{equation}
therefore relation (\ref{M2}) can be expressed as
\begin{equation} 
{\partial\gamma\over\partial n}={1\over 3}\pi^2k_B^2
{1+2\lambda\over 1+\lambda}
{\partial\ln{\cal D}_0\over\partial E_F}.
\label{M7}
\end{equation}

\subsection{Coefficient
$\partial\varepsilon_{\rm con}/\partial n$} 
The derivative of the condensation energy (\ref{EconTc})
includes the derivative of the critical temperature. For
Niobium and similar materials the critical temperature is
given by the McMillan formula,\cite{KSK74}
\begin{equation} 
T_c={\theta_D\over 1.45}\exp\left[-{1.04(1+\lambda)\over \lambda-
\mu^*(1+0.62\lambda)}\right],
\label{M8}
\end{equation}
where $\theta_D$ is the Debye temperature and $\mu^*$ is the
Coulomb pseudopotential. From (\ref{EconTc}) and (\ref{M8})
we express the condensation energy as
\begin{equation} 
\varepsilon_{\rm con}={\pi^2\over 12.6}k_B^2(1+\lambda){\cal D}_0
\theta_D^2\exp\left[-2{1.04(1+\lambda)\over\lambda-
\mu^*(1+0.62\lambda)}\right].
\label{M9}
\end{equation}

Experience from dilute alloys shows that the product ${\cal D}_0
\theta_D^2$ is nearly constant.\cite{VD76} We thus use as the
second approximation,
\begin{equation}  
{\partial\over\partial n}{\cal D}_0\theta_D^2=0.
\label{M10}
\end{equation}
In this approximation the derivative of the condensation energy
is given by the derivative of the factor $1+\lambda$ and by the
derivative of the argument of the exponential,
\begin{equation} 
{\partial\varepsilon_{\rm con}\over\partial n}=
\varepsilon_{\rm con}{\partial\over\partial n}
\left(-{2.08(1+\lambda)\over\lambda-\mu^*(1+0.62\lambda)}+
\ln(1+\lambda)\right).
\label{M11}
\end{equation}

Again, the experience with dilute alloys shows that the Coulomb
pseudopotential is nearly constant,\cite{VD76} therefore we take
as the third approximation,
\begin{equation} 
{\partial\mu^*\over\partial n}=0.
\label{M12}
\end{equation}
With approximation (\ref{M12}) the density derivative of the
condensation energy becomes proportional to the derivative of
the coupling parameter,
\begin{equation} 
{\partial\varepsilon_{\rm con}\over\partial n}=
\varepsilon_{\rm con}{\partial\lambda\over\partial n}
\left({2.08(1+0.38\mu^*)\over\left(\lambda-\mu^*(1+0.62\lambda)
\right)^2}+{1\over 1+\lambda}\right).
\label{M13}
\end{equation}

The density derivative of the coupling constant follows from
(\ref{M4}) and approximation (\ref{M5}) as
\begin{equation} 
{\partial\lambda\over\partial n}={V\over 2(1+\lambda)}
{\partial\ln{\cal D}_0\over\partial E_F}.
\label{M14}
\end{equation}
The derivative of the condensation energy is thus proportional to
the BCS interaction,
\begin{equation} 
{\partial\varepsilon_{\rm con}\over\partial n}=
{\varepsilon_{\rm con}V\over(1\!+\!\lambda)^2}
{\partial\ln{\cal D}_0\over\partial E_F}
\left({1.04(1\!+\!0.38\mu^*)(1\!+\!\lambda)\over
\left(\lambda-\mu^*(1+0.62\lambda)\right)^2}+{1\over 2}\right).
\label{M15}
\end{equation}

The material parameters for Niobium which we have used are listed
in Table~1. For convenience, we have included values which can
be evaluated from the above formulas, e.g., the critical
temperature is given by (\ref{M8}). The logarithmic derivative
of the density of states with respect to the energy is
extracted from the figure in Ref.~\onlinecite{EK77}. The hole
density $n$ has been evaluated from the London penetration
depth\cite{K76}
\begin{equation} 
\lambda_{\rm Lon}^2={m\over n_s e^2\mu_0}.
\label{M16}
\end{equation}
At zero temperature all holes are in the condensate,
$n=n_s$. The listed density of holes follows from
$\lambda_{\rm Lon}=\lambda_0=3.9\ 10^{-8}$~m and
the mass $m_0=1.2\,{\rm m}_e$. This effective mass is an
estimate of values 1.12, 1.6, 1.28 and 1.22 for different
orbits of the pure Niobium.\cite{BPK77}

We assume that the properties of the material are modified by
oxygen impurities of a concentration ranging from 0 to 0.03. We
neglect the effect of impurities on the thermodynamic parameters
taking into account only their dominant effect on the London
penetration depth and the GL coherence length. In the dirty
limit, the GL coherence length, defined in our model as
\begin{equation} 
\xi^2={n\hbar^2\over m^*\gamma(T_c^2-T^2)},
\label{M17}
\end{equation}
scales with the square-root of the mean free path $l$, $\xi\propto
\sqrt{l}$, while the effective London penetration depth scales
with its inverse, $\lambda_{\rm Lon}\propto 1/\sqrt{l}$.\cite{T96}
Accordingly, the GL parameter $\kappa=\lambda_{\rm Lon}/\xi$ is
proportional to the inverse mean free path, $\kappa\propto 1/l$.
One can see that the proper scaling of both characteristic lengths
is achieved by the scaling of the effective mass,
\begin{equation} 
m=m_0{\kappa_0\over\kappa_{\rm pure}},
\label{M18}
\end{equation}
where $\kappa_{\rm pure}$ is the GL parameter of the pure Niobium 
while $\kappa_0$ is the actual value for a given concentration of
impurities provided in Ref.~\onlinecite{KSK74}.

\begin{eqnarray}
&&
\begin{array}{llrl}
{\rm critical\ temperature}^{45}&T_c&9.5&{\rm K}\\
&&&\\
{\rm Debye\ temperature}^{45}&\theta_D&275&{\rm K}\\
&&&\\
{\rm coupling\ parameter}^{42}&\lambda&0.89&\\
&&&\\
{\rm Coulomb\ pseudopot.}^{42}&\mu^*&0.15&\\
&&&\\
{\rm coef.\ of\ spec.\ heat}^{45}&\gamma&719&
{\rm J}\,{\rm m}^{-3}{\rm K}^{-2}\\
&&&\\
{\rm mass\ in\ pure\ Nb}^{46}&m_0&1.2&m_e\\
&&&\\
{\rm hole\ density}^{45}&n&2.2\ 10^{28}&{\rm m}^{-3}\\
&&&\\
{\rm log.\ der.}^{43}&{\partial\ln{\cal D}_0\over\partial E_F}&
1.1\ 10^{19}&{\rm J}^{-1}\\
&&&\\
{\rm GL\ parameter}^{42}&\kappa_{\rm pure}&0.78&\\
&&&\\
{\rm density\ of\ states\ (A1)}&{\cal D}&5.7\ 10^{47}&
{\rm J}^{-1}{\rm m}^{-3}\\
&&&\\
{\rm bare\ density\ }\ldots\ {\rm (A3)}&{\cal D}_0&3.0\ 10^{47}&
{\rm J}^{-1}{\rm m}^{-3}\\
&&&\\
{\rm BCS\ interaction\ (A4)}&V&2.9\ 10^{-48}&{\rm J}\,{\rm m}^3\\
&&&\\
{\rm cond.\ energy\ (30)}&\varepsilon_{\rm con}&1.6\ 10^4&
{\rm J}\,{\rm m}^{-3}\\
&&&\\
{\rm cond.\ en.\ per\ pair}&{2\varepsilon_{\rm con}\over n}&
9.17\ 10^{-6}&{\rm eV}\\
&&&\\
{\rm coefficient\ (A7)}&
{1\over 2}{\partial\gamma\over\partial n}T_c^2
&3.85\ 10^{-6}&{\rm eV}\\
&&&\\
{\rm coefficient\ (A15)}&
{\partial\varepsilon_{\rm con}\over\partial n}
&8.73\ 10^{-6}&{\rm eV}\\
&&&\\
{\rm coefficient\ of}\ C_1&
{\partial\ln\varepsilon_{\rm con}\over\partial\ln n}
&1.9&\\
&&&\\
{\rm coefficient\ of}\ C_2&
{\partial\ln\gamma\over\partial\ln n}
&0.42&\\
&&&\\
\end{array}
\nonumber\\
&&{\rm Table~1.\ Material\ parameters\ of\ pure\ Niobium.}
\nonumber
\end{eqnarray}

\end{document}